\documentclass[twocolumn,showpacs,preprintnumbers,amsmath,amssymb,floatfix]{revtex4-1}
\usepackage{graphicx}
\usepackage{dcolumn}
\usepackage{bm}
\usepackage[all]{xy} 
\usepackage[latin1]{inputenc}
\usepackage{caption}
\usepackage{subcaption}
\usepackage{tikz}
\usepackage{comment}
\usetikzlibrary{shapes,arrows}
\usepackage{url}
\usepackage{pifont}

\newcommand{\ket}[1] {\left| #1 \right>}
\newcommand{\bra}[1] {\left< #1 \right|}
\newcommand{\braket}[2] {\left< #1 \middle| #2\right>}
\newcommand{\ketbra}[2] {\left| #1 \middle> \middle< #2\right|}
\newcommand{\braOpket}[3] {\left< #1 \left| #2 \right| #3 \right>}
\newcommand{\ketOpbra}[3] {\left| #1 \middle> #2 \middle< #3 \right|}

\newcommand{\pwf}[0] {\tilde\phi}
\newcommand{\seti}[1]{\left\{\ket{#1_i}\right\}}

\newcommand{\eref}[1]{Equation~\ref{eqn:#1}}
\newcommand{\sref}[1]{Section~\ref{sec:#1}}
\newcommand{\aref}[1]{Appendix~\ref{sec:#1}}
\newcommand{\fref}[1]{Figure~\ref{fig:#1}}

\newcommand{\tref}[1]{Table~\ref{tbl:#1}}

\graphicspath{{./figs/}}

\begin{document}

\title{The Shirley reduced basis: a reduced order model for plane-wave DFT}

\author{Maxwell Hutchinson}
\affiliation{The Physics Department, University of Chicago, Chicago IL 60637}
\email{maxhutch@uchicago.edu}
\author{David Prendergast}
\affiliation{The Molecular Foundry, Lawrence Berkeley National Laboratory, Berkeley CA 94720}

\date{\today}

\begin{abstract}
The Shirley reduced basis (SRB) represents the periodic parts of Bloch functions as linear combinations of eigenvectors taken from a coarse sample of the Brillouin zone, orthogonalized and reduced through proper orthogonal decomposition.
We describe a novel transformation of the self-consistent density functional theory eigenproblem from a plane-wave basis with ultra-soft pseudopotentials to the SRB that is independent of the k-point.
In particular, the number of operations over the space of plane-waves is independent of the number of k-points.
The parameter space of the transformation is explored and suitable defaults are proposed.
The SRB is shown to converge to the plane-wave solution.
For reduced dimensional systems, reductions in computational cost, compared to the plane-wave calculations, exceed 5x.
Performance on bulk systems improves by 1.67x in molecular dynamics-like contexts.
This robust technique is well-suited to efficient study of systems with stringent requirements on numerical accuracy related to subtle details in the electronic band structure, such as topological insulators, Dirac semi-metals, metal surfaces and nanostructures, and charge transfer at interfaces with any of these systems. 
The techniques used to achieve a k-independent transformation could be applied to other computationally expensive matrix elements, such as those found in density functional perturbation theory and many-body perturbation theory.

\end{abstract}

\maketitle

\section{Introduction} \label{sec:intro}
Electronic structure is a cornerstone of modern materials science and condensed matter research efforts.
Rapid advancements in experimental techniques have put pressure on the simulation community to provide efficient access to chemically accurate numerical predictions for a diverse range of materials, while high-throughput efforts~\cite{Jain2013} have embraced standard electronic structure methods, such as density functional theory (DFT)\cite{Kohn1965}, to explore entire databases of materials\cite{Belsky2002}.
For materials science applications, typically focused on condensed phases, surfaces, and extended nanostructures, the use of atomistic models employing periodic boundary conditions is the norm, with plane-wave DFT (PWDFT) providing a robust numerical approach for systems where the number of atoms does not extend to thousands. 
While PWDFT has been shown to be adequately robust in the treatment of exotic materials, whose electronic properties are defined by subtle details in electronic band structure (e.g., Dirac semi-metals and topological insulators~\cite{Wang2012}), numerically converged calculations of complex systems often come at prohibitive computational cost.
The Shirley reduced basis (SRB) technique is able to substantially reduce numerical effort while retaining the robustness of PWDFT.

PWDFT is a natural numerical representation for electronic structure under periodic boundary conditions, which are commonly employed to model crystalline and amorphous condensed phases, surfaces and their adsorbates, materials interfaces or heterojunctions, and extended nanostructures, such as sheets, ribbons, wires and tubes.
PWDFT can also be employed to study finite or molecular systems, but that will not be the focus of this work.
Note that finite temperature first-principles molecular dynamics simulations for condensed phases, surfaces, interfaces and extended nano structures typically also often rely on PWDFT, due to the robust description of electronic structure provided by this numerical representation, even for configurations far from equilibrium. 
In particular, we are interested in systems which are expensive to model due to the necessity to capture subtle details of electronic band structure which define the function of these materials.
Typically, this complexity is related to adequately describing the Fermi surface of complex metals or semi-metals, which requires detailed knowledge of electronic band structure with respect to electron wave vectors.

Bloch's theorem factors the representation of wave functions in periodic systems into a slowly oscillating phase, defined by the electron wave vector $k$, or \textit{k-point}, and a strictly periodic function, $u(r)$\cite{Martin}:
\begin{equation}
\psi(r) = e^{i k \cdot r} u(r) \qquad u(r+a) = u(r),
\end{equation}
where $a$ is a Bravais lattice vector of the periodic system.
In DFT, the Hamiltonian is block-diagonal with respect to the k-point, but the expectation value of the electron density relies on integrating contributions over all values of $k$ in the first Brillouin zone (BZ).
Numerical solutions to electronic structure problems discretely sample the BZ with a finite set of k-points, $k \in K \subset BZ$.
Existing electronic structure methods treat the k-dependent eigen-problems independently.
In reality, the eigenproblems and associated eigensolutions at each k-point contain a degree of redundancy.
The periodic Bloch states, $\ket{u_{nk}}$, are known to have relatively weak k-dependence in comparison to the dispersion relation, $\varepsilon_{nk}$, which is mainly driven by the quadratic k-dependence of the kinetic energy.

Shirley proposed that a reduced basis comprised of Bloch eigenstates over a coarse sample of the BZ be used to represent the eigenproblem throughout the BZ~\cite{Shirley1996}.
We denote the initial coarse sample as the \textit{q-points}.
Thus, the $n$th Bloch state at BZ sample $k$ can be represented as
\begin{equation}
\ket{u_{nk}} = \sum_{mq} \ket{u_{mq}} c^{nk}_{mq} .
\end{equation}
The number of q-points needed to accurately reproduce the solutions at all k-points is small.
We refer to this technique, the use of a coarse BZ sample as a basis, as \textit{Shirley interpolation}.
Analogies can be drawn to $k \cdot p$ theory~\cite{Luttinger1955a,Dresselhaus1955}, which can effectively interpolate electronic structure in the neighborhood of a specific k-point using a large number of eigenstates from that k-point alone.
By contrast, Shirley interpolation aims to provide a single reduced basis which spans the entire BZ by combining details from multiple q-points and a relatively small number of bands.

The basis size can be further reduced by selecting linear combinations of q-point states though a principal component analysis~\cite{Shirley1996}.
The principal components with small eigenvalues are removed from the basis, provided the specification of a tolerated error.
The use of principal component analysis in reducing the dimensionality of PDEs is often called proper orthogonal decomposition (POD)~\cite{Rathinam2003}.

The \textit{Shirley reduced basis} (SRB) is the union of Shirley interpolation and proper orthogonal decomposition.
Shirley applied the method to the non-self-consistent evaluation of band structure and associated spectroscopy and achieved speedups in excess of 1000$\times$ for dense k-point samples~\cite{Shirley1996}.

Prendergast and Louie extended Shirley's work to ultra-soft pseudopotentials (USPP) through interpolation of the projectors across the BZ~\cite{Prendergast2009}.
They also recognized that preserving periodicity with respect to $\bf k$ across the first BZ could be achieved using periodic images of the original coarse q-point basis.
In particular, the periodic images of the $\Gamma$-point were used to successfully enforce the periodic symmetry of the band structure.
Such images can be generated merely via mathematical transformation, without additional expensive plane-wave calculations.

Here, we further extend the method to self-consistent calculations, employing a new set of techniques to mitigate the added cost of constructing an electron density~\cite{Kohn1965}.
We also remove the interpolation of the atomic projectors required for the non-local potential, in favor of an auxiliary basis approach similar to the SRB for the periodic component of the Bloch eigenfunctions.
We implement the SRB in the popular PWDFT code Quantum Espresso~\cite{Giannozzi2009}.
Our implementation is freely available for general use~\footnote{\texttt{http://www.github.com/maxhutch/qe-srb}}.

The SRB approach is shown to be uniformly convergent and numerically efficient, achieving significant speedups for self-consistent calculations, including relaxations.
The accurate reproduction of forces and stresses also indicate the possibility for significant speed up of first-principles molecular dynamics simulations, which can prove prohibitively expensive for systems which require k-point sampling to capture metallic or semi-metallic behavior or the possibility of a transition to a metallic state via phase change~\cite{Pickard2011} or charge transfer~\cite{Chan2008}.
One can view such molecular dynamics simulations simplistically as requiring a converged self-consistent-field calculation at each time step for the determination of the next atomic configuration based on the computed forces and stresses.

In \sref{related}, we discuss similar methodologies and find the SRB to be the first combination of Shirley interpolation and proper orthogonal decomposition.
In \sref{algo}, we introduce the self-consistent extension of the SRB in general terms.
In \sref{examples}, we introduce structures that will be used to demonstrate the accuracy and performance of the SRB and describe our definition of the agreement between two calculations of the same physical structure.
In \sref{param}, we describe the parameters of the method and their affect on the solution.
In \sref{examples}, we describe representative physical systems that are used to demonstrate the effectiveness of the SRB and discuss agreement and convergence standards.
In \sref{conv}, we demonstrate the convergence properties and accuracy of SRB.
In \sref{perf}, we present a performance model for the SRB.
In \sref{disc}, we discuss interesting properties of the SRB.
In \sref{future}, we propose research efforts that would improve the efficiency and accuracy of the SRB.
Finally, in \sref{conc}, we conclude.
\aref{model} provides a more thorough performance model for the SRB algorithm.

\section{Related work} \label{sec:related}
The SRB is a reduced order method (ROM) based on Shirley interpolation and proper orthogonal decomposition (POD), which is related to principal component analysis and Galerkin projection in other fields~\cite{Lucia2004}.
POD, principal component analysis, and Galerkin projection have been well studied in fields as diverse as fluid dynamics~\cite{Sirovich1987}, machine learning\cite{Ding2004}, and image processing\cite{Rathinam2003,Homescu2005}.  
Shirley interpolation appears to have originated with the work of Shirley himself.

ROMs were developed in the fluid dynamics community~\cite{Sirovich1987,Epureanu2003} as a means to reduce the dimensionality of steady-state turbulent flows.
Samples of the velocity fluctuations $u - \langle u \rangle$ taken at intervals longer than the correlation time are fed into a singular-value decomposition to identify coherent structures in the flow.
The governing equations are then expressed in terms of the coherent structures, creating a model for the time-dependent energy transfer between the structures.
The SRB differs from common ROMs in two ways: 1) it models a non-linear system and 2) it samples in Bloch space rather than time.
The non-linearity makes a-priori error estimates difficult.
The Bloch-space sampling provides structure to the operator transformations.

The \textit{optimal representation of the polarization propogator} in \cite{Umari2009} uses the SVD of the product basis of Wannier-function expansions of wave functions to reduce the basis size, akin to POD. 
The representation does not, however, take advantage of the smoothness of a solution manifold as the SRB does in k-space and fluid dynamics ROMs do in time.
In other words, it lacks the interpolation step.
The optimal representation is most like the auxiliary basis used in this work, \sref{aux}, for the transformation of projectors and the reduced rank density matrix, neither of which involve k-space interpolation.

The $k \cdot p$-method has been applied to density functional theory calculations by Persson and Ambrosch-Draxl~\cite{Persson2007}.
It relies on analytic k-dependence of the Hamiltonian to express linear relations between the Bloch states at different k-points.
That reliance, however, restricts $k \cdot p$ to application to methods with only local potentials, such as FLPAW.
The SRB has no such restrictions, and is presented here with the generality of ultra-soft pseudo-potentials.

\textit{Reduced Bloch Mode Expansion} (RBME)~\cite{Hussein2006,Hussein2009} is an expansion in Bloch eigenfunctions nearly identical to the Shirley interpolation step, but lacking the POD step.
It has wide application beyond electronic structure to band structures of periodic systems and is agnostic to the discretization of the Bloch states.
The RBME technique post-dates Shirley's original work.

The ``reduced-basis'' method described by Pau~\cite{Pau2007} makes similar assumptions about the form of the potential in order to produce an eigenproblem that is affine in $k$.
This provides for an a posteriori error estimation. 
Like RBME, it lacks the POD step.

\section{Self-Consistent Algorithm} \label{sec:algo}
\begin{figure}
$$ \xymatrix@C=0em{
& \braOpket{g}{H[\rho]}{\psi_{mq}} &\overset{1}{=}& E_{mq} \braket{g}{\psi_{mq}} \ar@/^/[d] \\ 
& \ar@/_/[d]_3 \sigma^2 \braket{u_{mq}}{b}  &\overset{2}{=}& \braket{u_{mq}}{u_{m'q'}} \braket{u_{m'q'}}{b} \\ 
& \braOpket{b}{H[\rho](k)}{u_{nk}} &\overset{4}{=}& E_{nk} \braket{b}{u_{nk}} \ar@/^/[d]\\
&  \ar@/^5pc/[uuu]^0 \braOpket{r}{\rho}{r} &\overset{5}{=}& \sum_{nk} f_{nk}[E] \left|\braket{r}{\psi_{nk}}\right|^2
} 
$$
\caption{
Flow chart of the SRB method.
The primary self-consistent plane-wave method is embodied in steps 0, 1, and 5.
The SRB adds step 2 to build the reduced basis, step 3 to transform the k-dependent Hamiltonian into the reduced basis, and step 4 to solve the reduced eigenproblem.
}
\label{fig:flow}
\end{figure}
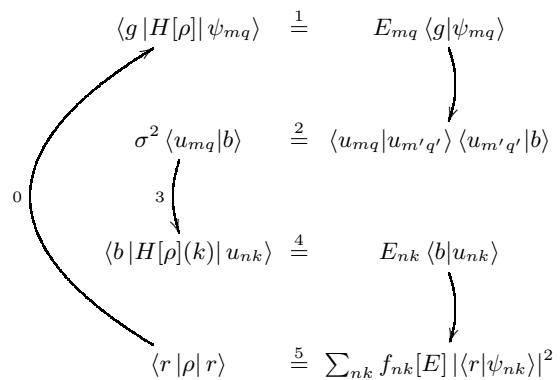

The methods of Shirley, Prendergast, and Louie for non-self-consistent interpolation are well described in the literature~\cite{Shirley1996,Prendergast2009}.
The most basic idea is to use the span of the periodic components of the Bloch eigenfunctions at a small number of points in the first BZ as a basis for the entire BZ.
Furthermore, a POD of these periodic functions allows for the basis size to be reduced.
Finally, a k-dependent Hamiltonian can be constructed without an explicit basis transformation through polynomial expansion of the kinetic energy and interpolation of projectors of the non-local potential.

The algorithm presented below is a natural generalization to the self-consistent regime.
It differs primarily in the addition of a step to reconstruct a real-space electron density from Bloch eigenfunctions in the reduced basis, and secondarily in more general approaches to the basis transformations and interpolations.

A few notes:  We use the term \textit{expansion} to refer to exact polynomial forms and reserve \textit{interpolation} for methods that include truncation errors.
The basis produced by this scheme is referred to as `reduced'\cite{Lucia2004}, as opposed to `optimal' in the previous literature, as it is only optimal within the space spanned by the periodic components of the input Bloch eigenfunctions.
For the sake of clarity, we assume the conventional method employs a plane-wave basis with pseudopotentials~\cite{Vanderbilt1990,Blochl1990}, though the method generalizes to any discretization of the Bloch eigen-functions, for example, in real-space using analytic or numerical atom-centered orbitals or using projector-augmented waves~\cite{Blochl1994}.
We use the unit cube in reciprocal lattice coordinates as our BZ, as opposed to the conventional choice of the origin-centered Wigner-Seitz cell.
 
The algorithm refines an electron density.
Generally, it would be repeated until a convergence criterion is met, signaling self-consistency.
The outline of the algorithm can be seen in Figure~\ref{fig:flow}: Start with an electron density represented in real-space.  
0) Represent the Hamiltonian as a functional of the electron density using a conventional basis, $\seti{g}$; 
1) Compute Bloch eigenfunctions $\braket{g}{\psi_{nq}}$ using a conventional method over a coarse reciprocal mesh, $Q$; 
2) Create a reduced basis $\seti{b}$ from an optimal subspace of the span of the periodic components of the conventional Bloch eigenfunctions; 
3) Represent k-dependent Hamiltonians in the new basis over a fine reciprocal mesh $K$; 
4) Solve the k-dependent Hamiltonians in the new basis producing the periodic components of the eigenfunctions $\braket{b}{u_{nk}}$; 
5) Construct a real-space electron density from the eigenfunctions in the new basis.

\subsection{Constructing the basis}\label{sec:basis}
The conventional algorithm produces eigenfunctions which are Bloch functions $\ket{\psi_{mq}} = e^{i q \cdot r} \ket{u_{mq}}$ on a reciprocal mesh $Q$.
Symmetries in the problem can be used to map the Bloch eigenfunctions computed at a single q-point to other points in reciprocal space.
In addition, the periodicity of eigensolutions in k-space permits states from the surface of the first BZ, e.g.\ the reciprocal space origin $q=\Gamma=0$, to be translated by any other reciprocal lattice vector, e.g.\ the corners of the unit cube, $0^3 \rightarrow \left\{0,1\right\}^3$ in reciprocal lattice coordinates.
This procedure is further described in~\cite{Prendergast2009}, and generally requires little numerical effort, given that such translations can be achieved merely by a reordering of the Fourier coefficients of each state.

Additional symmetries can provide more input q-points without additional plane-wave computation.
For example, inversion symmetry is computed by conjugating the plane-wave coefficients:
\begin{equation*}
c_{n,q} = c^*_{n,-q}.
\end{equation*}
which provides two interior k-points for the cost of one.
The symmetries can be compounded.
For example, a single q-point on the edge can be used to construct 7 other q-points without significant computation assuming inversion symmetry and periodicity:
\begin{multline*}
(0,0,1/4) \rightarrow \left\{(0,1,1/4), (1,0,1/4), (1,1,1/4), \right. \\\left. (0,0,3/4), (0,1,3/4), (1,0,3/4), (1,1,3/4)\right\}
\end{multline*}

We use POD to pick a subspace of the span of the periodic components of the Bloch eigenfunctions that is optimally representative.  First, construct a covariance (or overlap) matrix:
\begin{equation} \label{eqn:covar}
C_{i,j} \equiv \braket{u_i}{u_j} ,
\end{equation}
where $i,j$ are composite indices over $m,q$, the space of coarsely sampled eigen-functions.
Next, diagonalize the covariance matrix:
\begin{equation} \label{eqn:bdiag}
\sum_\gamma \braket{u_j}{u_\gamma} \braket{u_\gamma}{b_i} = \sigma_i^2 \braket{u_j}{b_i},
\end{equation}
where the eigenvalues $\sigma_i^2$ are the variances captured by the basis elements $\ket{b_i}$.  Select those basis elements, $\ket{b_i}$, with the largest variances as the reduced basis.
The size of the basis can be informed by choosing a sufficient number of basis elements to capture a significant fraction of the total variance. 
For example, the missing variance could be constrained to be below a certain fraction of the total variance:
\begin{equation}
1 - \frac{\sum_i^{N_b} \sigma_i^2}{\sum_j^N \sigma_j^2} < \sigma^2_{b}, 
\end{equation}
where $N_b$ is the number of basis elements, $N$ is the rank of the overlap matrix $C$, and $\sigma^2_b$ is the maximum tolerated error.

The basis elements are represented with respect to the original periodic components of the Bloch eigenfunctions, so one must generally transform them back to the original basis.
In the case of plane-waves, the transformation takes the form:
\begin{equation} \label{eqn:bexpand}
\braket{g_j}{b_i} = \sum_\gamma \braket{g_j}{u_\gamma} \braket{u_\gamma}{b_i} .
\end{equation}

\subsection{k-dependent Hamiltonians}
We use the reduced basis to expand a k-dependent Hamiltonian $H(k)=e^{-ik \cdot r} H e^{ik \cdot r}$.
In the plane-wave pseudopotential framework, the Hamiltonian has three components: the kinetic energy, the local potential, and the non-local potential~\cite{Martin,Pickett1989}:
\begin{equation}
H \equiv \ketOpbra{g}{T}{g} + \ketOpbra{r}{V}{r} + \ketOpbra{\beta}{V^{nl}}{\beta'}.
\end{equation}
It was previously shown that the kinetic energy could be written as a quadratic expansion in $\vec{k}$:
\begin{equation} \label{eqn:kin}
2 \braOpket{g_i}{T(\vec{k})}{g_i} = \left|\vec{k} + \vec{g}_i\right|^2 = |\vec{k}|^2 + 2 \vec{g_i} \cdot \vec{k} + |\vec{g}_i|^2 ,
\end{equation}
which can be transformed to the reduced basis by means of the matrices:
\begin{align} 
\braOpket{b_i}{T^0}{b_j} &= \sum_\gamma \braket{b_i}{g_\gamma} |\vec{g}_\gamma|^2 \braket{g_\gamma}{b_j} \label{eqn:kinsq} \\ 
\braOpket{b_i}{\vec{T}^1}{b_j} &= \sum_\gamma \braket{b_i}{g_\gamma} \vec{g} \braket{g_\gamma}{b_j} \label{eqn:kinlin} \\ 
\braOpket{b_i}{T^2}{b_j} &= \delta_{ij}=1 ,
\end{align}
such that
\begin{equation}
2 \braOpket{b_i}{T(\vec{k})}{b_j} = |\vec{k}|^2 T^2 + 2 \vec{k} \cdot \vec{T}^1 + T^0 ,
\end{equation}
which exactly isolates the $k$ dependence in the transformation of the kinetic energy.

The local potential is k-independent.
Its reduced basis representation is computed by explicit transformation from the plane-wave basis to real-space:
\begin{multline} \label{eqn:vloc}
\braOpket{b_i}{V^{loc}}{b_j} = \\ \sum_{\gamma,\gamma',r} \braket{b_i}{g_\gamma} \braket{g_\gamma}{r} V^{loc}(r) \braket{r}{g_{\gamma'}} \braket{g_{\gamma'}}{b_j} , 
\end{multline}
where the sums over $\gamma,\gamma'$ can be implemented as Fourier transformations.

The non-local potential is generally expressed as a matrix in the basis of atomic pseudo-wave functions, or just pseudo-functions: $\braOpket{\pwf_{a,l'}}{V^{nl}}{\pwf_{a,l}}$, where $a$ runs over atomic sites and $l$ over composite angular momentum quantum numbers~\cite{Kleinman1982}.
The non-local potential is diagonal for NCPP and block diagonal with respect to the atomic index for USPP.
We will occasionally compress the indices $(a,l) \rightarrow \alpha$.
The space of atomic wave functions is accessed through a dual-space of projectors, $\beta$, which couples isolated atomic Hilbert spaces to the overall periodic space comprising multiple atomic sites:
\begin{equation}
\sum_{l,l'} \ketbra{\beta_{a,l}}{\pwf_{a,l'}} = I_a, 
\end{equation}
where $I_a$ represents the identity in the space spanned by the pseudo-functions in a finite volume around atomic site $a$.
These matrix elements do not depend on $k$, but the projectors $\ket{\beta_{a,l}}$ do.
Transforming the projectors is all that must be done to transform the non-local potential to the reduced basis:
\begin{equation}
\braket{b_i}{\beta_\alpha(k)} = \sum_\gamma \braket{b_i}{g_\gamma} \braket{g_\gamma}{\beta_\alpha(k)}
\end{equation}
\begin{multline}\label{eqn:vnl}
\braOpket{b_i}{V^{nl}(k)}{b_j} = \\ \sum_{\alpha, \alpha'} \braket{b_i}{\beta_\alpha(k)} \braOpket{\pwf_\alpha}{V^{nl}}{\pwf_{\alpha'}} \braket{\beta_{\alpha'}(k)}{b_j}.
\end{multline}
For ultra-soft pseudo-potentials, the overlap matrix, $S$, is computed in the same way:
\begin{multline}
\braOpket{b_i}{S(k)}{b_j} - \delta_{i,j} = \\ \sum_{\alpha, \alpha'} \braket{b_i}{\beta_\alpha(k)} \braOpket{\pwf_\alpha}{S}{\pwf_{\alpha'}} \braket{\beta_{\alpha'}(k)}{b_j} .
\end{multline}
Note that when expressed in the basis of pseudo-functions, the non-trivial part of the overlap matrix, $\braOpket{\pwf_\alpha}{S}{\pwf_{\alpha'}}$, and non-local potential, $\braOpket{\pwf_\alpha}{V^{nl}}{\pwf_{\alpha'}}$, are often referred to as $Q$ and $D$, respectively~\cite{Vanderbilt1990}.

\subsection{Auxiliary basis for projectors} \label{sec:aux}
The projectors can be written as the product of origin-centered, k-dependent atomic projectors, $\beta_{0,l}(k)$, and an atom-dependent structure factor, $\mathcal{S}(a) = \exp[i(g+k)\cdot \tau_a]$, which we split into k-dependent and g-dependent terms:
\begin{equation} \label{eqn:proj}
\braket{g_i}{\beta_{a,l}(k)} = e^{- k \cdot \tau_a} \braOpket{g_i}{\mathcal{S}(a)}{g_i} \braket{g_i}{\beta_{0,l}(k)} ,
\end{equation}
where $l$ runs over angular momenta, $a$ runs over the atomic centers, and $a = 0$ corresponds to the origin.
The structure factor can be written as:
\begin{equation}
\braOpket{g}{\mathcal{S}(a)}{g} = e^{-i (g+k) \cdot \tau_a} ,
\end{equation}
If the number of k-points is large, we could consider transforming the structure factor in the SRB before applying it to the origin-centered projectors:
\begin{equation} \begin{aligned}
\braOpket{b_i}{\mathcal{S}(a)}{g_j} &= \braket{b_i}{g_j} \braOpket{g_j}{\mathcal{S}(a)}{g_j} \\
\braket{b_i}{\beta_{a,l}(k)} &= \sum_\gamma \braOpket{b_i}{\mathcal{S}(a)}{g_\gamma} \braket{g_\gamma}{\beta_{0,l}(k)}  .
\end{aligned} \end{equation}
The inner product in the plane-wave basis is inefficient: the size of the plane-wave basis is much larger than the rank of $\braOpket{b_i}{\mathcal{S}(a)}{g_j}$, which is the lesser of the size of the SRB and the dimension of the space spanned by the origin-centered projectors $\braket{g_\gamma}{\beta_{a,l}(k)}$.
We can introduce an auxiliary basis, $\ket{x}$, to reduce the effort in the inner product:
\begin{equation} \label{eqn:aux} \begin{aligned}
\braOpket{b_i}{\mathcal{S}(a)}{x_j} &= \sum_{\gamma} \braket{b_i}{g_\gamma} \braOpket{g_\gamma}{\mathcal{S}(a)}{g_\gamma} \braket{g_\gamma}{x_j} \\
\braket{x_i}{\beta_{0,l}(k)} &= \sum_\gamma \braket{x_i}{g_\gamma} \braket{g_\gamma}{\beta_{0,l}(k)} \\
\braket{b_i}{\beta_{a,l}(k)} &= \sum_j \braOpket{b_i}{\mathcal{S}(a)}{x_j} \braket{x_j}{\beta_{0,l}(k)}  .
\end{aligned} \end{equation}
The basis $x$ needs to represent the origin-centered projectors through the BZ.
This is analogous to the SRB, which must represent the periodic Bloch functions through the BZ.
Indeed, we use the procedure outlined in Section~\ref{sec:basis}, but with $\braket{g_i}{\beta_l(q,0)}$ as input states.
This auxiliary basis need only be produced once, as it is independent of the electron density and atomic positions.
Indeed, it could even be pre-computed and packaged with the pseudo-potential.
Therefore, we prefer to use the entire $K$ grid as the `coarse' sample $Q$, which makes the fractional variance a good metric for the accuracy of the basis.
In that sense, the auxiliary basis $\seti{x}$ is optimal.
This procedure must be performed for each atomic species.

\subsection{Diagonalizing the k-dependent Hamiltonian}
The Hamiltonian and, in the case of USPP, the overlap matrix are dense.
Therefore, it is natural to leverage highly optimized direct solvers, such as those found in the LAPACK library~\cite{Anderson1999}. However, the ratio of the number of basis elements to the number of bands is $O(10)$, which is large enough to justify the use of an iterative solver.
Furthermore, as in the conventional case, diagonalization in early self-consistent iterations need not be fully converged.
We intend to explore the use of iterative solvers in the future.

\subsection{Constructing the density matrix} \label{sec:rho}
After diagonalizing the k-dependent Hamiltonian to compute the periodic components of the eigenfunctions $\braket{b_i}{u_{mk}}$ and the energies $\varepsilon_{mk}$, one must produce a real-space electron density so the process can be repeated.
The simplest way of doing this would be to transform the eigenfunctions back to plane-waves and then accumulate the density in the usual way:
\begin{equation}
  \begin{aligned} \label{eqn:rsa}
  \braket{g_i}{u_{nk}} &= \sum_j \braket{g_i}{b_j} \braket{b_j}{u_{nk}} \\
  \braket{r}{u_{nk}} &= \sum_j \braket{r}{g_j} \braket{g_j}{u_{nk}}  \\
  \braOpket{r}{\rho}{r} &= \sum_{nk} \braket{r}{u_{nk}} f(\varepsilon_{nk})\braket{u_{nk}}{r}  ,
  \end{aligned}
\end{equation}
where $f(\varepsilon)$ is some occupation function and the transformation from plane-waves to real-space is generally accomplished with an FFT. 

Another method takes advantage of the small size of the SRB to form a density matrix:
\begin{equation}
\braOpket{b_i}{\rho}{b_j} = \sum_{nk} \braket{b_i}{u_{nk}} f(\varepsilon_{nk})\braket{u_{nk}}{b_j} .
\end{equation}
Using the Hermitian singular value decomposition $\rho = V \Sigma V^\dagger$, one can re-write the density matrix in terms of its rank-1 components:
\begin{equation}
\braOpket{b_i}{\rho}{b_j} = \sum_\nu \braket{b_i}{v_\nu} \sigma_\nu \braket{v_\nu}{b_j} , 
\end{equation}
where $\sigma_\nu$ are the diagonal elements of $\Sigma$.
The rank-1 components, $\braket{b_i}{v_\nu}$, can be accumulated in real-space just as we did for the periodic parts of the Bloch eigenfunctions in Equation~\ref{eqn:rsa}:
\begin{equation} \label{eqn:dma}
\begin{aligned}
\braket{g_i}{v_\nu} &= \sum_j \braket{g_i}{b_j} \braket{b_j}{v_\nu} \\
\braket{r}{v_\nu} &= \sum_j \braket{r}{g_j} \braket{g_j}{v_\nu} \\
\braOpket{r}{\rho}{r} &= \sum_\nu \braket{r}{v_\nu} \sigma_\nu \braket{v_\nu}{r}  .
\end{aligned}
\end{equation}
This method has the advantage that the number of rank-1 components is bounded above by the size of the SRB.
If there are more bands across all k-points than basis elements, $N_b N_k > N_r$, this approach performs fewer transformations.
In practice, many of the $\sigma_i$ are very small so they can be truncated just as small occupations $f(\varepsilon)$ generally are.
This can reduce the number of rank-1 components by up to a factor of 4.

In the case of USPP, there is an addition to the electron density:
\begin{multline}
\braOpket{r}{\tilde \rho}{r} = \\
\sum_{nk,\alpha,\alpha'} f(\varepsilon_{nk}) \braket{u_{nk}}{\beta_\alpha(k)} \braOpket{\pwf_\alpha}{S(r)}{\pwf_\alpha'} \braket{\beta_\alpha'(k)}{u_{nk}},
\end{multline}
which requires the inner product of the wave functions and the projectors:
\begin{equation} 
\braket{\beta_i(k)}{u_{nk}} = \sum_j \braket{\beta_i(k)}{b_j} \braket{b_j}{u_{nk}} .
\end{equation}
Note that the length of the inner product is the reduced basis size, representing a significant reduction in work over the plane-wave counterpart.

\subsection{Basis saving} \label{sec:basis_save}
The self consistent procedure serves not only to relax the electron density to the ground state, but also to relax the subspace spanned by the SRB to include the true ground state.  
Therefore, we consider two convergences: the convergence of the electron density and the convergence of the SRB.
To fully generalize the method, we consider these two convergences separately.
Namely, we can refrain from updating the SRB at every self-consistent iteration, instead `saving' the transformation matrix, kinetic energy, projectors, and factorized overlap matrix from the previous iteration.
The only parts of the eigenproblem which must be recomputed are the explicitly density dependent operators: the local and, in the case of USPP, non-local potentials, Equations~\ref{eqn:vloc} and~\ref{eqn:vnl}, respectively.
Relaxing the basis more slowly than the electron density can slow the convergence of the electron density, but the avoided cost of computing a new basis and transforming density independent quantities can result in a net performance improvement. 

The two convergences need not be terminated by the same error condition.
In general, the solution is less sensitive to the basis as it is to the electron density.
Thus, the basis can be frozen on a weaker convergence threshold than the electron density while preserving the accuracy of the solution.
In fact, if the basis is not frozen near convergence, it can lead to slowly decaying oscillations in the charge density and basis elements, similar to so-called charge sloshing, delaying convergence.

\subsection{Forces and stresses}
The electron density produced by the SRB technique, \eref{dma}, is represented in the same basis as the plane-wave result.
Therefore to compute forces and stresses, we need only address terms that depend on the wave-functions directly~\cite{Laasonen1993}.

The forces depend on derivative operators nominally expressed in the plane-wave basis:
\begin{equation}
\mathcal{D}^\mu(k) \equiv \braOpket{g_i}{\frac{\partial}{\partial R^\mu}}{g_j} = (g^\mu_i+k^\mu) \delta_{i,j} , 
\end{equation}
which are used to compute the derivatives of the projection:
\begin{equation}
\frac{\partial}{\partial R_a^\mu} \braket{\beta_{a,l}(k)}{u_{nk}} = \left[\bra{\beta_{0,l}(k)} \mathcal{S}(a)^\dagger \mathcal{D}^\mu(k)^\dagger \right] \ket{u_{nk}}.
\end{equation}
The matrices $\mathcal{S}$ and $\mathcal{D}$ are Hermitian and diagonal in the plane-wave basis, so we can drop the adjoints and commute them.
We need to compute the quantity:
\begin{equation} \label{eqn:dproj}
\braOpket{b_i}{\mathcal{S}(a) \mathcal{D}^\mu(k)}{\beta_{0,l}}, 
\end{equation}
so we can take inner products with the periodic Bloch states $\braket{b_i}{u_{nk}}$ in the reduced basis.

This problem is analogous to that of the projectors, \eref{proj}, but with $\mathcal{D}^\mu \beta_{0,l}$ replacing $\beta_{0,l}$.
Therefore, we can form an auxiliary basis just as in \eref{aux}.
As before, the auxiliary basis is independent of the electron density and the atomic position, so it can be produced once and stored for the duration of the calculation.
For relaxations and molecular dynamics, this allows the cost of producing the basis to be amortized by many force evaluations.
For self-consistent calculations with a single force evaluation, it may be advantageous to compute the derivatives of the projection in the plane-wave basis and explicity transform into the SRB.

The stress has an additional term that depends on the wave-functions, the kinetic stress~\cite{Nielsen1983}:
\begin{equation}
  \sigma^{\mu,\nu}_{\text{kin}} = \sum_{n,k,g} \braket{u_{nk}}{g} (g + k)^\mu (g+k)^\nu \braket{g}{u_{nk}}, 
\end{equation}
which leads to the k-dependent kinetic stress operator:
\begin{equation} \label{eqn:kstress}
 \braOpket{g}{\hat\sigma^{\mu,\nu}_{\text{kin}}(k)}{g} = (g + k)^\mu (g+k)^\nu . 
\end{equation}
Unsurprisingly, this term can be treated as the kinetic energy operator was in \eref{kin}: the product is expanded and the k-dependence factored yielding a k-independent transformation:
\begin{equation}
  \braOpket{g}{\hat\sigma^{\mu,\nu}_{\text{kin}}(k)}{g} = g^\mu g^\nu + g^\mu k^\nu + g^\nu k^\mu + k^\mu k^\nu, 
\end{equation}
the following nine transformations are required:
\begin{align}
&\braOpket{b_i}{g^\mu g^\nu}{b_j} &&= \sum_g \braket{b_i}{g} g^\mu g^\nu \braket{g}{b_j}\\
&\braOpket{b_i}{g^\mu}{b_j} &&= \sum_g \braket{b_i}{g} g^\mu \braket{g}{b_j}, 
\end{align}
which are then composed as:
\begin{multline}
  \braOpket{b_i}{\hat\sigma^{\mu,\nu}_{\text{kin}}(k)}{b_j} = k^\mu k^\nu \delta_{i,j} \\
+ \braOpket{b_i}{g^\mu g^\nu}{b_j}  + \braOpket{b_i}{g^\mu}{b_j} k^\nu + \braOpket{b_i}{g^\nu}{b_j} k^\mu  .
\end{multline}
Conveniently, the single-$g$ terms were already computed for the kinetic energy in \eref{kinlin}.
Further, one of the diagonal double-$g$ terms can be computed by subtraction:
\begin{equation}
g^3 g^3 = |g|^2 - g^1 g^1 - g^2 g^2, 
\end{equation}
with the $|g|^2$ term previously defined as $K^0$, \eref{kinsq}.
Thus, the total number of additional diagonal transformations is five.

The second derivatives found in the non-local contribution to the stress can be computed in a similar way as \eref{dproj}.
There are twelve such terms but they are density and atomic position independent, so they can be re-used.

\section{Examples} \label{sec:examples}

Through the rest of this paper, we discuss results from three physical systems: a $(3,3)$ carbon nanotube (CNT), a nickel slab decorated with water and hydroxide, and a bulk gold molecular dynamics snapshot at 2000 K.
Each system was chosen to be representative of a broad class of electronic structure problems, as described in the following sections.
To demonstrate the generality of the method, we use Perdew-Burke-Ernzerhof GGA (PBE)~\cite{Perdew1996} and Perdew-Zunger LDA (PZ)~\cite{Perdew1981} functionals with Vanderbilt ultra-soft~\cite{Vanderbilt1990}, Rappe-Rabe-Kaxiras-Joannopoulos ultrasoft (RRJK)~\cite{Rappe1990}, and Troulliers-Martins (TM)~\cite{Troullier1991} norm conserving pseudopotentials.
The raw input files used for these examples can be found in the \texttt{srb-supplemental} repository\footnote{\texttt{https://github.com/maxhutch/srb-supplemental}}.
A summaray of key parameters is given in~\tref{param}.

\subsection{(3,3) carbon nanotube}

\begin{figure}
\includegraphics[width=\columnwidth]{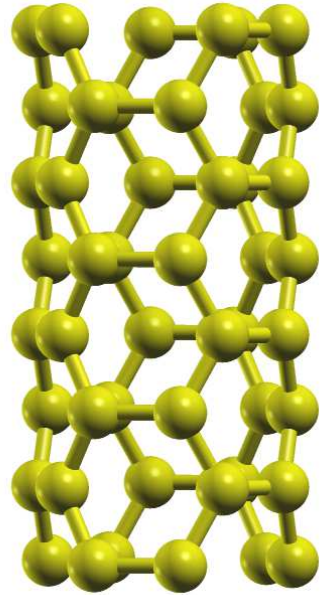}
\caption{Four cells of a (3,3) CNT.}
\end{figure}

The (3,3) carbon nanotube (CNT) is a metallic one-dimensional nano-structure.
It has a 12-atom primitive cell arranged around the z-axis.
To replicate isolation within periodic boundary conditions, a large vacuum regions is added in the x-y plane.
We are only concerned with relaxing stress along the tube axis.

Metallic systems with a small number of Fermi level crossings, such as the `Dirac points' in CNTs, require thorough sampling of the BZ.
When the BZ is under-sampled, the crossings can be missed, leading to an artificial band gap.
In this case, 32 irreducible points are required.

The CNT is representative of a large class of extended one-dimensional metallic and semiconducting molecules (e.g.\ linear polymers) and nanostructures,  including nanotubes, nanowires, and nanoribbons. 
Accurate descriptions of the electronic structure of these materials are vital to estimating their electronic properties and their structures. 
In addition, the proximity of metallic or semi-metallic nanostructures to molecules, nanoparticles or extended substrates often results in significant charge transfer across the interface due to electronic coupling and hybridization~\cite{Krepel2011}.
Inaccurate descriptions of the Fermi surface can modify the details of electron transfer and lead to poor descriptions of chemisorption or electronic transmittance across such interfaces.
Modeling such interactions at finite temperature would also require sufficient k-point sampling to cover the worst possibilities during an entire trajectory -- such as switching from metallic to semi-metallic behavior~\cite{Ganesh2011}.
In the absence of numerical convergence, such simulations would incorrectly describe the system as semiconducting in both cases.

\subsection{Nickel slab}

\begin{figure}
\includegraphics[width=\columnwidth]{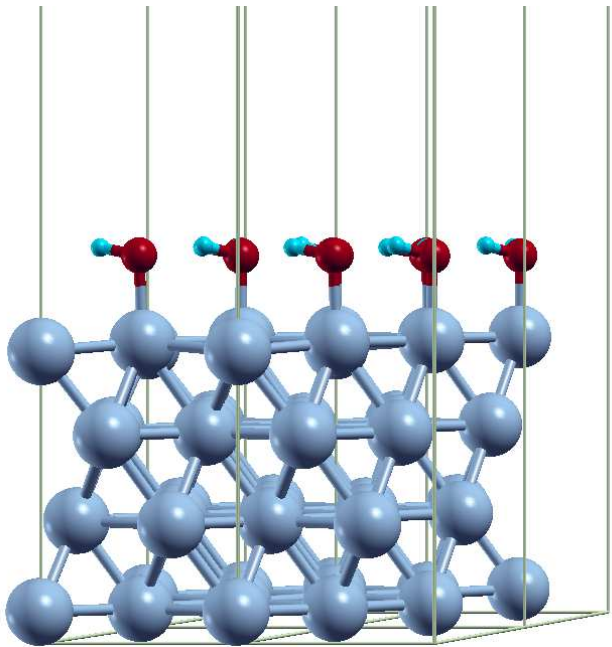}
\caption{Nickel slab decorated with water and hydroxide.  Cell is repeated $2 \times 2$ in the plane of the slab.}
\end{figure}

We provide an example relevant to the study of surface chemistry and catalysis. 
A four atomic layer Ni slab is chosen to model a semi-infinite reactive metal surface, with sufficient vacuum padding to reduce the influence of electronic coupling between the top and bottom surfaces. 
We decorate one side of the slab with water molecules and hydroxide moieties, representing a proposed surface coverage of Ni in the presence of water vapor~\cite{Li2010a}. 
The particular details of this system are not important within the context of the current study, and we will not try to draw any physical or chemical conclusions. 
However, this quasi-2D system is representative of a large class of simulations which aim to model surfaces and interfaces and their chemistry or reactivity.
Specific details of surface relaxation and reconstruction, charge transfer and reactivity, and electronic coupling via hybridization are all strongly dependent on an accurate description of the electronic structure and require numerically converged sampling of the BZ.
The combination of a large number of k-points and the additional requirement to include a large number of plane-waves to describe the vacuum above and below the slab render such common calculations prohibitively expensive. 
Any gains in efficiency through the use of the SRB would surely be welcome in the surface science and catalysis communities.

\subsection{Gold snapshot}
\begin{figure}
\includegraphics[width=\columnwidth]{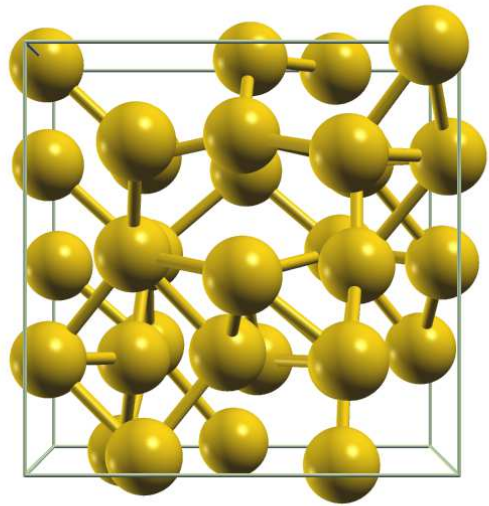}
\caption{Gold snapshot at 2000K.  Cell contains 32 atoms.}
\end{figure}

The gold system is a 32 atom disordered snapshot extracted from an MD simulation at 2000~K~\cite{Ping2008}.
A hard norm-conserving pseudo-potential is used to describe the ionic cores under these extreme conditions.
The gold snapshot is representative of the typical accuracy requirements and computational cost per step of first-principles molecular dynamics of metals under extreme conditions.
We wish to demonstrate the possibility of realizing some speedup for such systems.
In addition, we wish to provide a practicable option for running robust simulations of systems which undergo insulator to metal transitions or pressure induced metallization.
Furthermore, we hope to enable additional parallelization paradigms for large scale molecular dynamics simulations through our use of the SRB.

\subsection{Agreement} \label{sec:agree}

In the following sections, we will compare different calculations of the same physical system.
Here, we define the standards by which we compare calculations.
That is, what conditions need to be met for two calculations to be in `agreement'.

We define agreement in terms of three measures: the band structure ($\varepsilon$), the force ($F$), and pressure ($P$).  
Agreement in band structure means less than 5meV absolute root mean square error over bands that are below or cross the Fermi level.
We zero the band structure at the Fermi level to avoid constant shift errors.
Agreement in force means the root mean square error is less than $10^{-3} Ry/au$ or 5\% of the root mean square force, whichever is higher.
Agreement in pressure means the absolute error is less than 1 kbar or the relative error is less than 5\%.
The total energy is an interesting metric mathematically, as it converges monotonically with respect to basis completeness due to variational freedom.
However, it has no absolute physical meaning in pseudo-potential calculations.
Therefore, we report the total energy per atom ($E/A$), but do not constrain our methods to directly agree with respect to total energy.

\subsection{PWDFT convergence}

\begin{table}
\begin{ruledtabular}
\begin{tabular}{l | l r r r r r }
System & Calculation & $K$ & $R_v$ & $A$ & PP & $E_{xc}$ \\
\hline
CNT & SCF/Bands & 33 & $60 a_0$ & 12 & US & PBE \\
Ni  & SCF/Bands & 74 & $68 a_0$ & 17 & US & PBE \\
Au  & AIMD      & 36 & $ 0    $ & 32 & NC & PZ \\
\end{tabular}
\end{ruledtabular}
\caption{ \label{tbl:param}
Key parameters of representative examples.
$K$ is the number of irreducible k-points;
$R_v$ is cell dimension in the vacuum direction;
$A$ is the number of atoms;
PP is the pseudopotential type;
and
$E_{xc}$ is the exchange-correlation functional.
} 
\end{table}

PWDFT has two primary discretizations: the fine k-point mesh and the number of plane waves, which is defined through a kinetic energy cut-off $E_{cut}$.
Before comparing to the SRB, we ensure that that each structure is converged with respect to the number of k-points and plane-waves.
The resulting configurations can be found in \tref{param}.

For each structure, we converge with respect to the k-point mesh such that a doubling of the number of k-points on edge produces a result in agreement, as defined in \sref{agree}.
For example, if the converged mesh had $16$ points on edge, then a calculation with $32$ points on edge should agree with it.

Convergence with respect to energy cutoff is a more nuanced problem, especially for ultra-soft pseudo-potentials.
For demonstration purposes, we simplify the matter by using a 32 (48) Rydberg cutoff for ultra-soft (norm conserving) pseudo-potentials in our presented examples. 
The validity of the SRB over a broad range of energy cut-offs is demonstrated in \fref{ecut_conv}.

The CNT and nickel slab structures are vacuum padded normal to the structure to model a reduced dimensional system in full periodic boundary conditions.
At fixed atomic positions, the planar (axial) forces and stress of the Ni slab (CNT) are coupled to the amount of vacuum padding.
This prevents agreement with respect to the raw parallel pressure and forces at reasonable vacuum separations.
Instead, we allow each structure to relax normal to the vacuum direction before computing the agreement metrics.
However, for comparison with the SRB result, we use the un-relaxed with the vacuum from the relaxed structure.
This way, the forces and pressure are non-zero and can be meaningfully compared.

Band structure computations are performed non-self consistently with the same energy cut-off and vacuum padding as the self-consistent calculation.
However, in order to converge the band energies to high accuracy, sometimes the plane-wave eigensolver must be changed from Davidson to conjugate gradients.
The switch is not uncommon for non-self consistent calculations.

\section{Parameterization} \label{sec:param}
Thus far, we have tried to present the method as generally as possible.
Here, we describe the relevant parameters of the method and their effects on the accuracy and performance of the SRB.
In particular, we attempt to provide sensible defaults.

The times reported in this section should be interpreted qualitatively.
For quantitatively valid timings, see \sref{perf}.
 
\subsection{Defining the SRB}

\begin{figure}
\begin{subfigure}{\columnwidth}
\resizebox{\columnwidth}{!}{
\providecommand{\yTrans}[2]
{
    \pgftransformcm{0}{1}{0.4}{0.4}{\pgfpoint{#1cm}{#2cm}}
}

\providecommand{\yTransInverse}[2]
{
    \pgftransformcm{-1.}{2.5}{1}{0}{\pgfpoint{#1cm}{#2cm}}
}

\providecommand{\zTrans}[2]
{
    \pgftransformcm{1}{0}{0.4}{0.4}{\pgfpoint{#1cm}{#2cm}}
}

\providecommand{\zTransInverse}[2]
{
    \pgftransformcm{1}{0}{-1}{2.5}{\pgfpoint{#1cm}{#2cm}}
}

\begin{tikzpicture}

    \zTrans{0}{0};
    \draw [black!50,step=2cm] grid (8,0);

    \node at (0,0) [circle,fill=black] {};
    \node at (1,0) [circle,fill=black] {};
    \node at (2,0) [circle,fill=black] {};
    \node at (3,0) [circle,fill=black] {};
    \node at (4,0) [circle,fill=black] {};
    \node at (5,0) [circle,fill=white,draw] {};
    \node at (6,0) [circle,fill=white,draw] {};
    \node at (7,0) [circle,fill=white,draw] {};
    \node at (8,0) [circle,fill=white,draw] {};

    \node at (0,0) [outer sep=2pt,below]{(0,0,0)};
    \node at (8,0) [outer sep=2pt,below]{(1,0,0)};
\end{tikzpicture}}
\caption{\label{fig:BZ1}
Stick
}
\end{subfigure}
\begin{subfigure}{\columnwidth}
\resizebox{\columnwidth}{!}{
\providecommand{\yTrans}[2]
{
    \pgftransformcm{0}{1}{0.4}{0.4}{\pgfpoint{#1cm}{#2cm}}
}

\providecommand{\yTransInverse}[2]
{
    \pgftransformcm{-1.}{2.5}{1}{0}{\pgfpoint{#1cm}{#2cm}}
}

\providecommand{\zTrans}[2]
{
    \pgftransformcm{1}{0}{0.4}{0.4}{\pgfpoint{#1cm}{#2cm}}
}

\providecommand{\zTransInverse}[2]
{
    \pgftransformcm{1}{0}{-1}{2.5}{\pgfpoint{#1cm}{#2cm}}
}

\begin{tikzpicture}

    \zTrans{0}{0};
    \draw [black!50,step=2cm] grid (8,8);

    \node at (0,0) [circle,fill=black] {};
    \node at (4,4) [circle,fill=black] {};
    \node at (2,0) [circle,fill=black] {};
    \node at (4,0) [circle,fill=black] {};
    \node at (6,0) [circle,fill=white,draw] {};
    \node at (8,0) [circle,fill=white,draw] {};
    \node at (0,2) [circle,fill=black] {};
    \node at (0,4) [circle,fill=black] {};
    \node at (0,6) [circle,fill=white,draw] {};
    \node at (0,8) [circle,fill=white,draw] {};

    \node at (8,2) [circle,fill=white,draw] {};
    \node at (8,4) [circle,fill=white,draw] {};
    \node at (8,6) [circle,fill=white,draw] {};
    \node at (8,8) [circle,fill=white,draw] {};
    \node at (2,8) [circle,fill=white,draw] {};
    \node at (4,8) [circle,fill=white,draw] {};
    \node at (6,8) [circle,fill=white,draw] {};

    \node at (0,0) [outer sep=2pt,below]{(0,0,0)};
    \node at (8,0) [outer sep=2pt,below]{(1,0,0)};
    \node at (0,8) [outer sep=2pt,left]{(0,1,0)};

\end{tikzpicture}}
\caption{\label{fig:BZ2}
Slab
}
\end{subfigure}
\begin{subfigure}{\columnwidth}
\resizebox{\columnwidth}{!}{
\providecommand{\yTrans}[2]
{
    \pgftransformcm{0}{1}{0.4}{0.4}{\pgfpoint{#1cm}{#2cm}}
}

\providecommand{\yTransInverse}[2]
{
    \pgftransformcm{-1.}{2.5}{1}{0}{\pgfpoint{#1cm}{#2cm}}
}

\providecommand{\zTrans}[2]
{
    \pgftransformcm{1}{0}{0.4}{0.4}{\pgfpoint{#1cm}{#2cm}}
}

\providecommand{\zTransInverse}[2]
{
    \pgftransformcm{1}{0}{-1}{2.5}{\pgfpoint{#1cm}{#2cm}}
}

\begin{tikzpicture}

    \draw [black!50,step=4cm] grid (8,8);

\begin{scope}
    \yTrans{8}{0};
    \draw [black!50,step=4cm] grid (8,8);
\end{scope}
\begin{scope}
    \yTrans{0}{0};
    \draw [black!50,dashed,step=4cm] grid (8,8);
\end{scope}
\begin{scope}
    \zTrans{0}{8};
    \draw [black!50,step=4cm] grid (8,8);
\end{scope}
\begin{scope}
    \zTrans{0}{0};
    \zTransInverse{0}{8};
    \draw [black!50,dashed,step=4cm] grid (8,8);
\end{scope}
\begin{scope}
    \zTrans{0}{0};
    \draw [black!50,dashed,step=4cm] grid (8,8);

    \node at (0,0) [circle,fill=black] {};
    \node at (4,0) [circle,fill=black] {};
    \node at (0,4) [circle,fill=black] {};
    \node at (4,4) [circle,fill=black] {};
    \node at (8,0) [circle,fill=white,draw] {};
    \node at (0,8) [circle,fill=white,draw] {};
    \node at (4,8) [circle,fill=white,draw] {};
    \node at (8,4) [circle,fill=white,draw] {};
    \node at (8,8) [circle,fill=white,draw] {};

    \node at (0,0) [outer sep=2pt,below]{(0,0,0)};
    \node at (8,0) [outer sep=2pt,below]{(1,0,0)};
    \node at (0,8) [outer sep=2pt,left]{(0,1,0)};
\end{scope}
\begin{scope}
    \zTrans{0}{8};
    \node at (0,0) [circle,fill=white,draw] {};
    \node at (4,0) [circle,fill=white,draw] {};
    \node at (0,4) [circle,fill=white,draw] {};
    \node at (4,4) [circle,fill=white,draw] {};
    \node at (8,0) [circle,fill=white,draw] {};
    \node at (0,8) [circle,fill=white,draw] {};
    \node at (4,8) [circle,fill=white,draw] {};
    \node at (8,4) [circle,fill=white,draw] {};
    \node at (8,8) [circle,fill=white,draw] {};
    \node at (0,0) [outer sep=2pt,left]{(0,0,1)};
\end{scope}
\begin{scope}
    \zTrans{0}{4};
    \node at (0,0) [circle,fill=black] {};
    \node at (4,0) [circle,fill=black] {};
    \node at (0,4) [circle,fill=black] {};
    \node at (4,4) [circle,fill=black] {};
    \node at (8,0) [circle,fill=white,draw] {};
    \node at (0,8) [circle,fill=white,draw] {};
    \node at (4,8) [circle,fill=white,draw] {};
    \node at (8,4) [circle,fill=white,draw] {};
    \node at (8,8) [circle,fill=white,draw] {};
\end{scope}
\end{tikzpicture}}
\caption{\label{fig:BZ3}
Bulk.
}
\end{subfigure}
\caption{ \label{fig:BZ}
Coarse samples of the first Brillouin zone.
Open points are related to closed points by inversion and translation.
}
\end{figure}
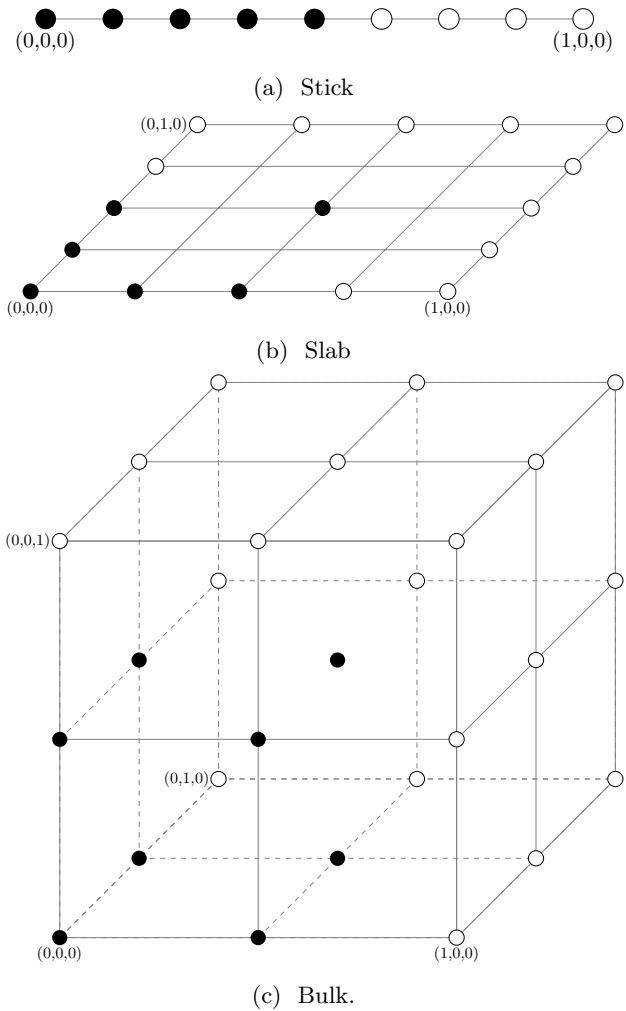

\begin{table}
\begin{ruledtabular}
\begin{tabular}{l | r r | 
D{.}{.}{2.2} D{.}{.}{2.3} D{.}{.}{1.6} D{.}{.}{3.2}  
}
& $N_Q$ & $N_b$ & 
\multicolumn{1}{c}{$\Delta E/A$} & \multicolumn{1}{c}{$\Delta \varepsilon$} & \multicolumn{1}{c}{$\Delta F$} & \multicolumn{1}{c}{$\Delta P$} \\
\hline
Corners & 8 & 1595 & 31.74 & 7.27 & 0.003617 & 173.26 \\
Edges &  20 & 3458 &  1.24 & 0.39 & 0.000231 &   1.29 \\
Faces &  26 & 3507 &  1.03 & 0.21 & 0.000092 &   0.55 \\
Center & 27 & 3509 &  1.00 & 0.20 & 0.000082 &   0.51 \\
\end{tabular}
\end{ruledtabular}
\caption{ \label{tbl:qpoints}
Comparison of basis size and accuracy for Au with various set of q-points.
In all cases, $\sigma^2_b = 10^{-7}$.
Energies are given in $meV$, forces in $Ry/a_0$, and pressure in kbar.
}
\end{table}

\begin{table}
\begin{ruledtabular}
\begin{tabular}{D{.}{.}{2}  r | 
D{.}{.}{2.2} D{.}{.}{2.3} D{.}{.}{1.7} D{.}{.}{3.2} 
} 
\multicolumn{1}{c}{$\sigma^2_b$} & $N_b$ & 
\multicolumn{1}{c}{$\Delta E$} & \multicolumn{1}{c}{$\Delta \varepsilon$} & \multicolumn{1}{c}{$\Delta F$} & \multicolumn{1}{c}{$\Delta P$} \\
\hline
10.^{-4}  & 2173 & 99.62 & 13.640 & 0.002883 & 92.17 \\
10.^{-7}  & 3507 &  1.03 &  0.209 & 0.000092 &  0.55 \\
10.^{-14} & 5485 &  0.50 &  0.043 & 0.000038 &  0.47 \\
0.        & 5486 &  0.50 &  0.043 & 0.000038 &  0.47 \\
\end{tabular}
\caption{ \label{tbl:trace_tol}
Comparison of basis size and accuracy Au with various error tolerances, $\sigma^2_b$.
In all cases, the q-points are corners, edges, and faces of the unit cube.
Energies are given in $meV$, forces in $Ry/a_0$, and pressure in kbar.
}
\end{ruledtabular}
\end{table}

There are three parameters that govern the subspace of the full basis in which the reduced basis is generated: the coarse sample of reciprocal space, Q, the truncation threshold, $\sigma^2_b$, and the maximum number of basis elements, $N_{max}$.  
Our implementation provides for a simultaneous definition of $N_{max}$ and $\sigma^2_b$.  
As SCF calculations converge, the number of basis elements that satisfy $\sigma^2_b$ tends to decrease, so if both are specified tightly, $N_{max}$ applies early in the calculation and $\sigma^2_b$ later.

It was found that for non-self consistent calculations of sufficiently large unit cells, the $\Gamma$ point and its mirrors on the unit cube are generally sufficient~\cite{Prendergast2009}.
For self-consistent calculations, this is not generally the case.
We find it best to use points that lie on the boundary of the BZ.
For example, in three dimensions the corners, edge centers, and face centers of the unit cube should be used in that order.
The relative basis sizes and accuracy for those coarse samples can be seem in \tref{qpoints}.
In two dimensions, only the face $q_z = 0$ of the unit cube should be included and the coarse spacing should be halved.
In one dimension, only the edge $q_z = q_y = 0$ should be included and the spacing halved again. 
These coarse BZ samples can be seen in \fref{BZ}.

On could consider increasing the number of input coarse eigenstates by artificially increasing the number of bands per coarse sample beyond the number desired in the fine sample.
However, the additional high-energy eigenstates would do little to improve the SRBs representation of the occupied low-energy eigenstates.
Simply, the high and low energy eigen-spaces overlap less than the low energy eigenspaces across the BZ.

This is indicative of a more general trade-off: the more representative the coarse sample $Q$ is of the fine sample $K$, the fewer basis elements are needed to achieve a fixed error tolerance.
For a fixed error tolerance, selecting a more representative $Q$ generally leads to a smaller basis.
This is more readily understood by holding the number of basis elements fixed and increasing the density of $Q$, which will reduce the error.
In addition to the aforementioned example of increasing the number of bands, this principal applies to applying symmetry beyond the first BZ.

The relationship between the basis truncation threshold, $\sigma^2_b$, the basis size, and the accuracy is demonstrated in \tref{trace_tol}.
To achieve agreement, we recommend $\sigma^2_b = 10^{-7}$ as a default.

\subsection{Auxiliary basis}

\begin{table}
\begin{ruledtabular}
\begin{tabular}{l r | 
D{.}{.}{2.2} D{.}{.}{2.3} D{.}{.}{1.6} D{.}{.}{3.2}
}
$\sigma^2_x$ & \% Elem. & 
\multicolumn{1}{c}{$\Delta E$} & \multicolumn{1}{c}{$\Delta \varepsilon$} & \multicolumn{1}{c}{$\Delta F$} & \multicolumn{1}{c}{$\Delta P$} \\
\hline
$10^{-2}$ & 10.24 & 51.59  & 73.75  & 0.000224 & 3.74 \\
$10^{-3}$ & 56.94 &  1.36  &  0.685 & 0.000093 & 0.54 \\
$10^{-4}$ & 84.72 &  1.08  &  0.211 & 0.000092 & 0.55 \\
$10^{-6}$ & 100.  &  1.03  &  0.209 & 0.000092 & 0.55 \\
0         & 100.  &  1.03  &  0.209 & 0.000092 & 0.55 \\
\end{tabular}
\end{ruledtabular}
\caption{\label{tbl:aux_tol}
Comparison of the number of auxillary basis elements, as a fraction of the total projector input states, and accuracy for Au with various auxillary error tolerances.
In all cases, the auxillary q-point grid $Q = K$.
Energies are given in $meV$, forces in $Ry/a_0$, and pressure in kbar.
}
\end{table}

\begin{table}
\begin{ruledtabular}
\begin{tabular}{l | D{.}{.}{3.2} D{.}{.}{3.2}  }
System & \multicolumn{1}{c}{Direct (s)} & \multicolumn{1}{c}{Auxillary basis (s)} \\
\hline
CNT &   47.98 &   20.59 \\
Ni  & 3009.32 & 2612.15 \\
Au  &  104.73 &   77.34 \\
\end{tabular}
\end{ruledtabular}
\caption{\label{tbl:aux_perf}
Comparison of the run-time of the direct transformation and auxiliary basis approaches to the projectors.
In all cases, $\sigma^2_x = 0.001$.
}
\end{table}

The auxiliary basis $\ket{x}$ used in transforming the projectors is controlled by three parameters (\sref{aux}) in the same manner as the SRB.
Because the auxiliary basis need be constructed only once, but is used at every iteration, it is advantageous to be thorough in the sample $Q$. 
We let $Q = K$, which gives the variances of the basis elements statistical meaning and produces an `optimal' basis.
In this light, the threshold method is preferred over specifying the number of elements.
A comparison of the auxiliary basis size and resulting accuracy for various thresholds is provided in \tref{aux_tol}.
A threshold $\sigma^2_x = 10^{-3}$ is a robust choice.

The dimension of the SVD used to form the auxiliary basis scales linearly with the number of k-points.
For small systems with very large number of k-points, this cost can dominate the overall cost of projection.
In those cases, a smaller set of q-points, $Q \subset K$, should be used.

The construction of the auxiliary basis for each species is independent of the number of atoms of that species while the construction of the structure factors occurs once per atom.
The overhead associated with constructing the auxiliary basis is amortized by the number of atoms.
If the number of atoms of a species is small, the overhead costs are more of a factor.
Therefore, it may only be beneficial to use the auxiliary basis when the number of atoms of a species is above some threshold, $N_{x,min}$.

Similarly, for non-self-consistent calculations there is only one projection.
Thus, the overhead cost of the SVD is not amortized over many iterations.
For non-self-consistent calculations, direct transformation of the projectors is recommended.

The three examples here are self-consistent with a moderate number of k-points.
Thus, we expect the auxiliary basis to provide superior performance.
It does, as seen in \tref{aux_perf}.

\subsection{Density matrix}

\begin{table}
\begin{ruledtabular}
\begin{tabular}{l D{.}{.}{3.1} | 
D{.}{.}{2.2} D{.}{.}{2.3} D{.}{.}{1.6} D{.}{.}{3.2}
}
$\sigma^2_\rho$ & \multicolumn{1}{c|}{\% $N_b$} & 
\multicolumn{1}{c}{$\Delta E$} & \multicolumn{1}{c}{$\Delta \varepsilon$} & \multicolumn{1}{c}{$\Delta F$} & \multicolumn{1}{c}{$\Delta P$} \\
\hline
$10^{-2}$ &  23.4 & 6.42 & 11.93  & 0.000564 & 0.13 \\
$10^{-3}$ &  29.6 & 8.07 &  3.79  & 0.000568 & 0.03 \\
$10^{-4}$ &  37.0 & 8.11 &  2.636 & 0.000580 & 0.01 \\
$10^{-6}$ &  48.2 & 8.11 &  2.635 & 0.000582 & 0.01 \\
0         & 100.  & 8.11 &  2.620 & 0.000582 & 0.01 \\
\hline
n/a         & \multicolumn{1}{c|}{n/a}  & 8.11 & 2.600 & 0.000537 & 0.01 \\
\end{tabular}
\end{ruledtabular}
\caption{ \label{tbl:rho_tol}
Comparison of the number of rank-1 components, as a fraction of the basis size, and accuracy for CNT with various density error tolerances.
The last entry uses the direct transformation.
Energies are given in $meV$, forces in $Ry/a_0$, and pressure in kbar.
}
\end{table}

\begin{table}
\begin{ruledtabular}
\begin{tabular}{l | D{.}{.}{3.2} D{.}{.}{3.2}  }
System & \multicolumn{1}{c}{Direct (s)} & \multicolumn{1}{c}{Density matrix (s)}  \\
\hline
CNT &  511.75 &   47.68 \\
Ni  & 2958.32 &  588.99 \\
Au  & 3976.52 & 2223.18 \\
\end{tabular}
\end{ruledtabular}
\caption{ \label{tbl:rho_perf}
Comparison of the run time of the direct transformation and density approaches to computing the real-space electron density.
In all cases, the density error tolerance $\sigma^2_\rho = 10^{-4}$.
}
\end{table}

There are two parameters that describe the building of the real-space density.
The first is the threshold on the Fermi weight with which to add a wave function to the density matrix.
This threshold is present in most PW DFT codes.
The second is the threshold on the singular values of the density matrix, $\sigma_\rho$.
Because the density matrix is complete, this parameter provides an exact measure for the preservation of the electron density.
The relationship between this threshold and accuracy is demonstrated in \tref{rho_tol}.
A sensible value is $10^{-4}$.

The density matrix is the size of the Hamiltonian, so the rank-1 decomposition costs no more than a single k-point diagonalization.
The performance of this approach is compared to the direct transformation in \tref{rho_perf}.

\subsection{Basis saving}

\begin{table}
\begin{ruledtabular}
\begin{tabular}{l | r r | 
D{.}{.}{2.2} D{.}{.}{2.3} D{.}{.}{1.6} D{.}{.}{3.2}
}
$n_L$ & $N_I$ & Time (s) & 
\multicolumn{1}{c}{$\Delta E$} & \multicolumn{1}{c}{$\Delta \varepsilon$} & \multicolumn{1}{c}{$\Delta F$} & \multicolumn{1}{c}{$\Delta P$} \\
\hline
5 & 52 & 1039.51 & -3.879 & 2.308 & 0.000746 & -0.35 \\
4 & 47 & 1044.89 & -3.879 & 2.303 & 0.000745 & -0.35 \\
3 & 56 & 1306.05 & -3.879 & 2.299 & 0.000746 & -0.34 \\
2 & 54 & 1425.08 & -3.879 & 2.300 & 0.000746 & -0.34 \\
1 & 52 & 1883.34 & -3.879 & 2.296 & 0.000747 & -0.33 \\
\end{tabular}
\end{ruledtabular}
\caption{\label{tbl:basis_life}
Comparison of the number of iterations, run-time, and accuracy for Ni with various basis life-times.
In all cases, the freeze threshold $\epsilon_b = 10^{-6}$.
Energies are given in $meV$, forces in $Ry/a_0$, and pressure in kbar.
}
\end{table}

\begin{table}
\begin{ruledtabular}
\begin{tabular}{l | r r | 
D{.}{.}{2.2} D{.}{.}{2.3} D{.}{.}{1.6} D{.}{.}{3.2}
}
$\epsilon_b$ & $N_I$ & Time (s) &
\multicolumn{1}{c}{$\Delta E$} & \multicolumn{1}{c}{$\Delta \varepsilon$} & \multicolumn{1}{c}{$\Delta F$} & \multicolumn{1}{c}{$\Delta P$} \\
\hline
$10^{-2}$ & 47 &  941.19 & -1.7085 & 2.260 & 0.000851 & -0.38 \\
$10^{-3}$ & 47 & 1052.83 & -1.6878 & 2.199 & 0.000932 & -0.31 \\
$10^{-4}$ & 49 & 1189.98 & -1.6861 & 2.197 & 0.000944 & -0.20 \\
$10^{-6}$ & 50 & 1612.63 & -1.6847 & 2.191 & 0.000947 & -0.29 \\
$10^{-9}$ & 50 & 2163.92 & -1.6850 & 2.206 & 0.000949 & -0.29 \\
0         & 50 & 2347.25 & -1.6878 & 2.199 & 0.000932 & -0.31 \\
\end{tabular}
\end{ruledtabular}
\caption{ \label{tbl:freeze}
Comparison of the number of iterations, run-time, and accuracy for Ni with various freeze thresholds.
In all cases, the basis lifetime $n_L = 1$.
Energies are given in $meV$, forces in $Ry/a_0$, and pressure in kbar.
}
\end{table}

There are two parameters that define the reuse of the SRB during self-consistent calculations: the basis lifetime, $n_L$, and the freeze threshold, $\epsilon_b$.
The basis lifetime is the minimum number of SCF iterations for which each basis is reused.
The freeze threshold is the energy convergence criterion after which the basis is no longer updated.

As shown in \tref{basis_life}, the accuracy is nearly completely independent of the basis lifetime.
As the basis lifetime increases, the number of iterations to convergence, $N_I$, can increase.
However, the mitigation of overhead costs generally makes up for any additional iterations, lowering overall runtime.
Across the full range of systems, $n_L = 3$ is a robust choice.

As the wave functions and electron density converge, the SRB changes less and less between updates.
When the energy eigenvalues are converged to $\epsilon_b$, the standard convergence criteria for self-consistent calculations, the SRB is not updated.  
As demonstrated in \tref{freeze}, the accuracy rapidly improves as the threshold is decreased until basis saving is no longer a source of error.
$\epsilon_b = 10^{-6}$ is a conservative choice.

\section{Accuracy} \label{sec:conv}
\begin{figure}
\centering
\resizebox{\columnwidth}{!}{\input{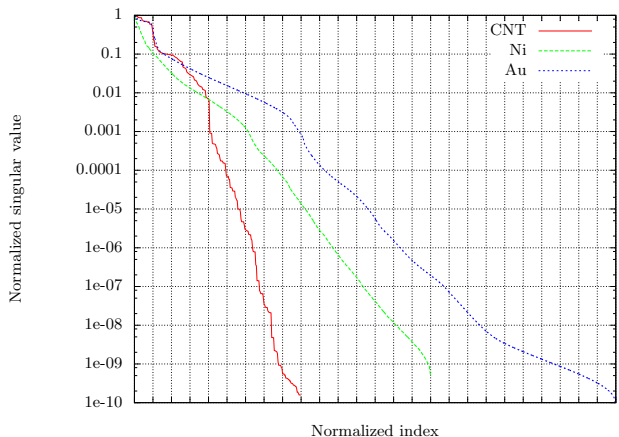}}
\caption{\label{fig:vars}
Singular value spectrum taken from the end of a self-consistent calculation using the coarse samples specified in \fref{BZ}.
The vertical axis is the singular value divided by the number of coarse samples, that is the number of q-points $Q$.
The horizontal axis is the singular value index divided by the number of eigenfunctions taken per q-point, that is the number of bands $B$.
}
\end{figure}

\begin{figure*}
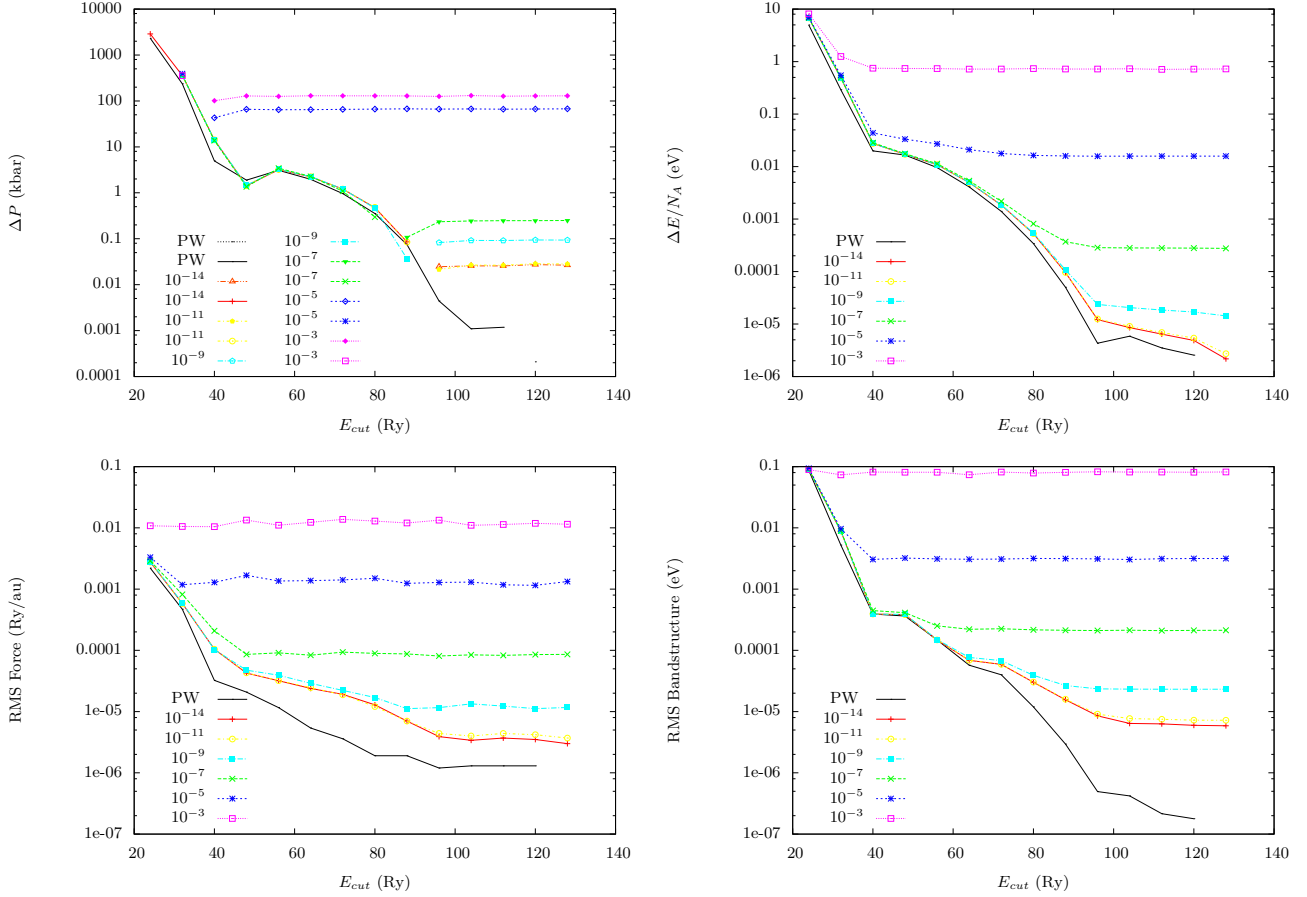

\centering
\begin{subfigure}{\columnwidth}
\resizebox{\columnwidth}{!}{\input{figs/pp_conv_P.tex}}
\end{subfigure}
\begin{subfigure}{\columnwidth}
\resizebox{\columnwidth}{!}{\input{figs/pp_conv_E.tex}}
\end{subfigure}
\begin{subfigure}{\columnwidth}
\resizebox{\columnwidth}{!}{\input{figs/pp_conv_F.tex}}
\end{subfigure}
\begin{subfigure}{\columnwidth}
\resizebox{\columnwidth}{!}{\input{figs/pp_conv_B.tex}}
\end{subfigure}
\caption{ \label{fig:ecut_conv} 
Convergence of the total energy, band structure, forces, and pressure with respect to $E_{cut}$ compared to $E_{cut} = 128$ Ry for the bulk Au example.
The black lines are for a PW calculation and the colored lines for SRB calculations with a range of basis truncation thresholds, $\sigma_b^2$.
In the error in pressure is signed, so positive errors are dashed lines.
}
\end{figure*}

There is an important inconsistency between the plane-wave basis used to construct the SRB and the plane-wave basis used in a plane-wave calculation.
Plane-wave codes typically define their plane-wave bases as the set of g-vectors within a sphere centered around each k-point.
This ensures that each k-point has the same well-defined $E_{cut}$, but each k-point therefore uses a different set of g-vectors.
\begin{equation}
G_{PW}(k) = \left\{g : |k+g|^2 < E_{cut}\right\}
\end{equation}
The SRB, on the other hand, is k-independent; we typically define the SRB elements with respect to the $\Gamma$-point g-vectors.
Similar issues arise when the mirroring procedure is applied to the input q-points~\cite{Prendergast2009}, taking $q \rightarrow q'$.
In the plane-wave code, the plane-wave basis at $q$ and $q'$ would be different. 

We would like to create a larger basis that includes every k-point basis by simply adding a g-vector in each direction.
That is, a 28x34x60 reciprocal space grid would become a 30x36x62 grid.
However, in most modern PWDFT codes, including Quantum ESPRESSO, a custom FFT interface is used which imposes an isotropic energy cut-off. 
We could increase the $E_{cut}$ such that every g-vector at every plane-wave k-point is included in the $\Gamma$-point basis.
Each SRB k-point would then be represented on a superset of g-vectors from a mix of plane-wave k-points.
\begin{equation}
G_{SRB} = \left\{ g : |g|^2 < E_{cut} + |k_{max}|^2 \right\} \supset \bigcup_{k \in K} G_{PW}(k)
\end{equation}

If the original PW calculation is not fully converged with respect to the plane-wave cutoff, then the SRB with increased cut-off will converge to a different, but sometimes more accurate result, due to the basis discrepancy.
However, the differences induced by basis discrepancy closely track the g-vector truncation error.
The error of the PW and SRB results, computed with respect to a fully g-vector converged result, are nearly identical until the intrinsic accuracy defined by the basis truncation threshold is reached.
Beyond that point, the SRB error levels off.
The position of this level off is the true error of the SRB.
This effect is demonstrated in Figure~\ref{fig:ecut_conv}.
By decreasing $\sigma_b^2$, the error can be driven well below agreement standards defined in \sref{agree}.

\subsection{Convergence}

The SRB spans a subspace of the span of the PW basis.
As the SRB grows, by adding samples to the coarse grid or truncating at lower error threshold, it converges uniformly to the PW result.
Furthermore, convergence of variational quantities, such as the total energy, is monotonic.

The basis elements are produced by singular value decomposition, Eq.~\ref{eqn:bdiag}.
Singular value spectra of the three model systems can be seen in Figure~\ref{fig:vars}.
The steady exponential fall-off of the singular values demonstrates the redundancy of the set of input states $\{\ket{u_{mq}}\}$, for $q$ sampled from $Q$, and suggests that the solution should converge exponentially with respect to the basis truncation threshold $\sigma^2_b$.

Convergence of numerical examples displays three regimes.  
First, when the SRB is too small, the non-linearity of the system can cause a high-error plateau region.
Subsequently, the error falls off exponentially.
Finally, other differences between the cutoffs become apparent, causing the SRB difference to level off at the scale of the PW cutoff convergence.

There are other model differences related to the size of the auxiliary basis and the thresholds used for truncating sums in the density matrix.
These errors are negligible when using the parameters described in Section~\ref{sec:param}.

\section{Performance} \label{sec:perf}
\begin{table*}
\begin{ruledtabular}
\begin{tabular}{l | r r r | r r r | r r r}
    & PW-0   & PW-K   & PW Total & SRB-0  & SRB-K  & SRB Total & $K_S$ & $X_\infty$ & $X_{real}$ \\
\hline
CNT & 196.06 & 91.13 & 3204. & 470.25 &  3.99 & 602. & 4 & 22.84 & 5.32 \\
Ni & 234.95 & 181.69 & 13680. & 2551.6 &  52.28 & 6420. & 18 & 3.47 & 2.13 \\
Au & 10.93 & 341.36 & 12300. & 6783.55 & 246.57 & 15660. & 72 & 1.38 & 0.78 \\
\end{tabular}
\end{ruledtabular}
\caption{ \label{tbl:k_perf}
Breakdown into k-independent and k-dependent costs.
PW-K (SRB-K) is the time spent in k-depdenent steps of the PW (SRB) calculation divided by the number of k-points.
PW-0 (SRB-0) is the time spent in k-indepdenent steps of the PW (SRB) calculation.
$K_S$ is the number of k-points needed for the SRB require less time than the PW calculation.
$X_\infty$ is the ratio of the PW to SRB calculation times in the limit as $K \rightarrow \infty$.
$X_{real}$ is the ratio of the PW to SRB calculation times for the actual calculation.
}

\end{table*}

\begin{table}
\begin{ruledtabular}
\begin{tabular}{l | r r r r r r}
& K & Q & G & S & $\alpha_D$ & $\alpha_S$  \\
\hline
CNT & 33 & 5 & 79214 &  176 & 2.54 & 6.07\\
Ni  & 74 & 7 & 12431 &  972 & 2.02 & 11.85 \\
Au  & 36 & 7 & 20672 & 3016 & 2.39 & 14.29 \\
\end{tabular}
\end{ruledtabular}
\caption{ \label{tbl:indicators}
Indicators of the relative performance of the SRB.
K is the number of irreducible k-points;
Q is the number of irreducible q-points;
G is the number of plane waves;
S is the number of SRB elements;
$\alpha_D$ is the size of the subspace used in the Davidson algorithm as a ratio of the number of bands;
and
$\alpha_S$ is the size of the SRB as a ratio of the number of bands.
}
\end{table}

\begin{table}
\begin{ruledtabular}
\begin{tabular}{l | r r r r r r r}
& PW & $\{\ket{b}\}$ & $H$ & $H(k)$ & Diag & $\rho$ \\
\hline
CNT & 62.6 &  4.7 & 19.9 &  6.4 &  0.7 & 5.8 \\
Ni  & 16.0 &  2.6 & 12.8 & 17.2 & 43.8 & 7.7 \\
Au  & 12.9 & 11.2 & 14.6 &  5.2 & 46.8 & 9.3 \\
\end{tabular}
\end{ruledtabular}
\caption{ \label{tbl:srb_prof}
Percent of run-time spent in plane-wave, basis construction, k-independent transformation, k-dependent construction, diagonalization, and density matrix sections, respectively.
}

\end{table}

\begin{table}
\begin{ruledtabular}
\begin{tabular}{l | r r r r r}
    & $\sigma^2_b$ & $\sigma^2_x$ & $\sigma^2_\rho$ & $n_L$ & $\epsilon_b$ \\
\hline
CNT & $10.^{-6}$ & 0          & $10.^{-4}$ & 1 & $10.^{-2}$ \\
Ni  & $10.^{-5}$ & $10.^{-3}$ & $10.^{-4}$ & 5 & $10.^{-5}$ \\
Au  & $10.^{-6}$ & $10.^{-3}$ & $10.^{-4}$ & 4 & $10.^{-4}$ \\
\end{tabular}
\end{ruledtabular}
\caption{ \label{tbl:opt_param}
Optimal parameters for example systems such as to acheive agreement.
}

\end{table}

A simple empirical performance analysis is to break down the computation time into a setup and a per k-point cost, as seen in \tref{k_perf}.
PWDFT spends the overwhelming majority of time in k-dependent terms, with only a few $\rho$-dependent operations.
The SRB, on the other hand, spends a significant amount of time in k-independent `overhead', in the form of the q-point calculations and basis transformations.
The overhead is balanced by significantly lower per k-point costs.
For the examples we have considered, 
the number of k-points needed to amortize the overhead, resulting in an overall faster calculation, ranges from 4 to 72,
the speedup in the limit of an infinite number of k-points ranges from 1.4 to 22x,
and
the speedup of real calculations ranges from 0.8 to 5.3x.

It should be noted that the results presented in \tref{k_perf} were optimized for minimal total time.
The optimal parameters for these specific systems can be see in \tref{opt_param}.
Greater per k-point acceleration could be achieved at the cost of greater overhead, and vice versa.

The SRB computation can be divided into the q-point plane-wave solutions, the construction of the SRB $\{\ket{b}\}$, the k-independent transformation of the Hamiltonian, the construction of the k-dependent Hamiltonian, diagonalization, and the construction of the electron density.
The time spent in these sections of the code can be seen in \tref{srb_prof}.

For CNT, the SRB calculation is dominated by the plane-wave parts, which includes the q-point solutions and the k-independent transformation of the local potential.
Diagonalization, on the other hand, takes less than 1\% of the run-time.
In this case the SRB can be thought of as a way to reduce the number of FFTs, which dominate the vacuum-padded calculation.

For Ni and Au, the focus shifts from plane-wave calculations to dense linear algebra.
Au spends more time building the basis because bulk BZs have more q-points than slab BZs.
Ni spends more time constructing the k-dependent Hamiltonian and overlap matrix, $S$, because it employs ultra-soft pseudo-potentials.
In both cases, diagonalization is a significant cost, but no individual cost dominates the calculation.
The cost of each SRB step is at least linear in the number of basis elements, so reducing the basis size would reduce the cost of these calculations.

We can also reason about the performance of the SRB analytically.
A full performance model, based on counting library calls, can be found in \aref{model}.
The model highlights three parameters as strongly indicative of the performance of the SRB compared to plane-waves:
1) the k-point ratio, $Q / K$;
2) the band density, $B / V$;
and 
3) the subspace inefficiency, $\alpha_S / \alpha_D$.
The first two parameters are known a priori, while the third is buried deeply in the method of diagonalization.
The value of these parameters for the three examples systems are found in \tref{indicators}.
In all cases, lower values favor the SRB.

The k-point ratio, $Q/K$, characterizes the balance of over-head to per k-point costs.
The lower the value, the less significant the overhead of the q-point plane-wave calculations.
The k-point ratio captures the importance of the number of k-points on the relative efficiency of the SRB.

The band density provides an a priori estimate of the ratio of the SRB size to the number of plane-waves.
Plane-waves cover space uniformly; the number of plane-waves is linearly dependent on the volume of the unit cell.
The SRB is computed from the bands at each q-point.
Thus, the size of the SRB depends linearly on the number of bands, not the volume.
The ratio of the SRB size to the number of planes thus goes as the ratio of the number of bands to the volume, the band density. 
The band density captures the importance of the contents of the unit cell on the relative efficiency of the SRB.

The subspace inefficiency is defined as the ratio of the SRB size to the size of the iterative subspace that is directly solved in the iterative eigen-solver.
Because the SRB currently uses a direct solver for diagonalization, this ratio measures the relative difficulty of direct diagonalization in the two methods.
The cost of diagonalization is cubic in the dimension, so the relative costs are the cube of the subspace inefficiency.
The subspace inefficiency captures the difficulty of spanning the entire BZ with a single basis, compared to computing an iterative subspace for each k-point individually.
In that context, it is clear why the subspace inefficiency is greater than one.

Unlike the k-point ratio and band density, the subspace inefficiency can not be computed a priori.
We have found it to depend strongly on the dimensionality of the BZ.
The subspace inefficiency is highest for the 3D BZ of bulk systems and lowest for the 1D BZ of nanowire, nanoribbons, and nanorods.

The data presented thus far reflects serial performance as a proxy for model complexity.
In reality, PWDFT problems are solved in parallel, motivating a comparison of parallel performance.
Such a comparison requires considerations of the parallel model employed both in the SRB and PW calculations, which are not entirely consistent and beyond the scope of this writing.
However, a qualitative discussion of the benefits of the SRB algorithm for parallel performance is discussed in \sref{parallel}.

\subsection{Tuning} \label{sec:tune}

\begin{table}
\input{dat/tune.tbl}
\end{table}

The performance results presented thus far come from parameter configurations that satisfy all three agreement conditions from \sref{agree}.
In some cases, these agreement conditions can be relaxed.
For example, if the geometry is prescribed, then force and stress agreement for the purpose of relaxing the structure is unnecessary.
On the other hand, if the end result is a molecular dynamics trajectory, then a highly accurate band structure may not be required at every time-step.
The SRB provides a means for expressing lessened accuracy expectations to yield higher throughput calculations.

Here, we explore two situations with reduced accuracy requirements.
The first is AIMD, where the band-structure and stress at every time step are not important.
We further relax the force constraint to be within 10\% of the root mean square magnitude.
The second is a band structure calculation of a prescribed CNT, where the forces and stress are not important.
The constraint on the root mean square error of the band structure is kept at 5 meV.

The relaxed accuracy constraints lead to improved performance, seen in \tref{tune}.
The resulting SRB performance for is 1.67x better than plane-waves for Au and 5.89x better for CNT.

\section{Discussion} \label{sec:disc}

\subsection{Where to expect speedup}

The performance model is able to capture the major indicators of SRB performance: the ratio of k-points to q-points, the ratio of plane-waves to reduced basis elements, and the ratio of reduced basis elements to bands.

Three dimensional bulk materials in modest cells often require a large number of k-points to resolve the Fermi surface.
When disordered, broken symmetries further increase the number of irreducible points.
However, bulk materials also have relatively high band densities.
Furthermore, the entire volume of the BZ must be covered by the basis, requiring more basis elements per band than in a system with a 1D or 2D BZ.
Using direct diagonalization, as the current implementation does, the SRB will provide no more than a modest reduction in the run-time.
Using iterative diagonalization, one could reduce the performance model's dependence on the ratio of basis elements to bands, providing a speedup through more efficient application of the Hamiltonian.

One dimensional wire, tube, or ribbon materials require vacuum padding to imitate isolation in periodic boundary conditions.
The padding increases the number of plane-waves without changing the number of occupied bands, that is it decreases the band density.
Also, the BZ is one-dimensional, so relatively few basis elements per band are needed.
However, a linear BZ generally requires fewer k-points.
One dimensional systems frequently approach the limit given by \eref{klimit} with $Q \approx 4$ and $\alpha_S \approx 2 \alpha_D$.

Slab geometries are the best of both worlds.
The need for vacuum padding decreases the band density compared to bulk systems.
The two planar dimensions can require many k-points compared to the one-dimensional case.
The k-points lie in a plane, which can be spanned my a smaller number of basis functions than the bulk.
The modest performance of the Ni slab considered here can be attributed to the thickness of the slab and lack of iterative diagonalization.

\subsection{Implications for parallelism} \label{sec:parallel}
The SRB adds global operations to the SCF process: the construction of the basis and the transformation of k-independent factors of the Hamiltonian.
However, these operations contain inner products over the plane-wave space, which is large and readily parallelizeable.
Only the diagonalization of the covariance matrix, Eq.~\ref{eqn:bdiag}, exhibits poor scaling, and that operation is both G and K independent.

A significant advantage of the SRB comes in the reduction of the problem size.
The reduced, dense eigenvalue problem is significantly smaller than the plane-wave equivalent.
For example, a system with 1000 bands and 20 basis elements per band would fit on a 16 GiB node.
Generally, the SRB requires less memory and therefore fewer nodes to handle each k-point.
This allows for aggressive k-point parallelism, which further improves scalability through memory and network locality local.

The operations required to build the k-independent parts of the Hamiltonian have memory requirements that scale with the number of plane-waves.
The transformation matrix, $\braket{b}{g}$, can be particularly large.
However, the application of these matrices is through highly parallel matrix-matrix products.
Therefore, a scheme which parallelizes the construction of a set of dense Hamiltonians over a set of nodes and then subdivides into maximally local solves parallelizes very well compared to the plane-wave counterpart.

Additionally, the SRB uses significantly fewer FFT operations than the plane-wave counterpart.  
Parallel FFTs have high communication to computation ratios and are known to limit plane-wave DFT scalability in many cases.
Avoiding the majority of them eases this parallel bottleneck.

\subsection{Implications for accelerators}
Accelerator systems, including those based on GPUs and coprocessors, can struggle to provide sufficient network bandwidth to keep up with their high floating-point and local memory performance.
The aforementioned decrease in problem size, increase in possible locality, and avoidance of parallel FFTs reduces the communication overhead significantly.
The dense linear algebra that comprises the majority the SRB computation is an ideal case of accelerators.
Early testing indicates that the potential for the application of accelerators to the SRB exceeds that of traditional PW calculations.

\subsection{Advantages for norm-conserving pseduo-potentials}
The SRB presents three advantages for norm-conserving pseudo-potentials.
First, the norm-conserving non-local potential is density independent, and therefore static throughout a calculation.  
When the SRB is saved, only the local potential must be recomputed between SCF iterations.
Second, the eigensystem does not need to be generalized which greatly improves the efficiency of direct solvers compared to subspace methods that still require generalized solves.
Third, the usual disadvantage of NCPPs is the requirement of more plane-waves than USPPs.
The performance of the SRB is relatively insensitive to the number of plane-waves compared to full plane-wave calculations.
We can expect the SRB to reduce the performance gap between NCPPs and USPPs. 

\subsection{Comparison to NSCF calculations}

There are three primary differences between self-consistent and non-self-consistent calculations in the SRB.
The first, and most obvious, is the addition of a step to construct the charge density.
The method outlined in \sref{rho} is usually cheaper than the transformation of the local potential, so this should not significantly impact the performance of the method.

The second difference is the reduction in the number of k-points.
The density does not depend as strongly on the k-point sampling as the band-structure.
It is not uncommon for NSCF calculations to exceed thousands of k-points.
It is rare for self-consistent calculations to reach one hundred.
This shifts the target away from creating the smallest, most expressive basis towards methods that facilitate cheap transformations, reducing the overhead cost.
This shift is typified by the basis saving technique.

The third difference is related to the diagonalization.
Iterative diagonalization schemes take as input a convergence threshold for the accuracy of the eigenvalues.
The first self-consistent iteration specifies a large threshold.
Subsequent iterations use successively smaller thresholds until the user-defined accuracy criteria is met.
Because each SCF iteration uses the previous iterations result as an `initial guess', the diagonalizer has very little work to do at each iteration.
In a sense, self-consistent calculations only diagonalize once, but they interleave that diagonalization with self-consistent updates to the Hamiltonian.
In NSCF calculations, the same iterative diagonalizer is used but the convergence threshold can not be relaxed.
Therefore, NSCF diagonalization takes much longer than a single SCF iteration.

The SRB uses direct diagonalization of the dense Hamiltonian, which doesn't benefit much from a reduced convergence threshold.
The SRB diagonalization time in the SCF and NSCF cases are identical.
Thus, the reduction in diagonalization time compared to plane-waves is much more pronounced for NSCF calculations, even if the number of k-points, plane-waves, and basis elements were identical.

\section{Future work} \label{sec:future}

\subsection{Preconditioned iterative diagonalization}

We have seen the direct diagonalization step to be a significant bottleneck for higher dimensional systems.
The poor performance is rooted in the larger size of the SRB compared to the subspace formed in the Davidson iteration employed by the plane wave code.
Direct diagonalization is cubic in the matrix dimension, so even a doubling of the SRB size compared to the subspace will degrade performance by 8x.

The obvious solution is to use an iterative method, such as Davidson, to diagonalize the SRB Hamiltonian.
This would allow the SRB to take advantage of gradual diagonalization inherit to self-consistent calculations.
The iterative subspace needed to diagonalize the SRB Hamiltonian should be no larger than the PWDFT counterpart, and could indeed be smaller.

The challenge, as highlighted by Shirley~\cite{Shirley1996}, is to pick a reasonable preconditioner.
The SRB Hamiltonian is dense and not always diagonally dominant.
The SRB provides a factorization of the Hamiltonian with respect to $k$.
One could consider directly inverting the k-independent part of the Hamiltonian and re-using it as the foundation of dense preconditioner.
Such a preconditioner could be much more accurate than the diagonal preconditioners used in the plane-wave code, enabled by the small dense matrix size.

\subsection{Pruning}
The success of the basis saving technique hints at a temporal-like redundancy in the self-consistent procedure.
One could take this temporal redundancy a step farther and not only reuse a previous basis but reduce it further based on statistics of the previous iteration.
For example, on the first iteration, the reduced basis could be produced with little or no truncation.
Then, at the end of the iteration, the weight of the states $\braket{b_i}{u_{nk}}$ on each basis element would indicate which elements to remove.
This would amount to computing scores
\begin{equation}
\omega_i = \sum_{nk} f(\epsilon_{nk}) \braket{b_i}{u_{nk}}
\end{equation}
and removing some of the elements with the lowest score.
This could be up to a sum of missing scores or a ratio of the number of elements.
We call this pruning.
Pruning could be repeated at the end of each iteration until the basis is frozen or a new, un-truncated basis could be formed.

There are two advantages of pruning over simply producing a new basis.
First, pruning takes into consideration the Bloch states at all the k-points, not just the q-points.
Therefore, we could expect pruning to better decide which basis elements are being used.
Secondly, pruning does not require a new transformation of the density-independent components of the Hamiltonian.
Instead, the pruned Hamiltonian is formed by eliminating rows and columns from SRB Hamiltonian and rows from the set of projectors.

\subsection{Dividing the BZ}

In the present scheme, the entire first BZ is covered by a single reduced basis.
This requirement is the primary contributor to subspace inefficiency.
In Davidson, the subspace need only be valid for a single k-point.

The BZ could be divided into regions, each covered by a smaller basis.
For example, the first BZ could be split into octants, the first octant covering $[0,1/2]^3$ in crystal coordinates.
We would expect that at a given error tolerance, the octant reduced bases would be smaller than the basis for the entire BZ.
Dividing the BZ in this way would increase the overhead transformation costs, but reduce the per k-point cost.
If the number of k-points were very large, the reduction in per-k-point cost due to the smaller reduced bases could outweigh the added overhead cost.

\subsection{Exact exchange}
The computation of the exact-exchange operator in the reduced basis presents many challenges.
The exchange operator can be written as
\begin{equation}
\braOpket{b_i}{K(k)}{b_j} =\sum_{nk} f_{nk}(\varepsilon) \frac{\braket{b_i}{r} \braket{r}{\psi_{nk}} \braket{\psi_{nk}}{r'} \braket{r'}{b_j}}{ | r - r'| }, 
\end{equation}
up to constant prefactors.
This expression is generally treated by considering a two-particle density:
\begin{equation}
\rho_{a,b}(r) = \braket{\phi_a}{r} \braket{r}{\phi_b}, 
\end{equation}
where $\phi_a$ and $\phi_b$ are arbitrary complex fields (e.g.\ the wave function $\psi_{nk}$ or basis element $b_j$).
Then an analogous two-particle potential is formed through a Poisson solve.
Both of these steps are defined in specific bases, real space for the density and reciprocal space for the Poisson solve.
As linear operators, the two operators are a 2,2 tensor, which is too large to be direct transformed or stored in the SRB.

Another approach would be to recognize that the sum over occupied states produces a non-diagonal density operator:
\begin{equation}
\braOpket{r}{\rho}{r'} = \sum_{nk}f_{nk}(\varepsilon) \braket{r}{\psi_{nk}} \braket{\psi_{nk}}{r'}.
\end{equation}
We would like to perform a rank-1 decomposition, as in \sref{rho}, but the off-diagonal elements of the density do not benefit from cancellation of the k-point phase.
That is, the non-diagonal density operator is full rank.

However, it is not the opinion of the authors that the exact exchange operator can not be expressed in the SRB.
Simply, this is an area which requires further study.

\section{Conclusions} \label{sec:conc}

We have shown the SRB to be capable of achieving an arbitrary degree of accuracy compared to the host calculation in the primary basis, in this case plane-waves.
More generally, the SRB completes to the primary basis and therefore inherits the accuracy and convergence properties of that basis.
For plane-waves, this makes the SRB uniformly convergent.

The current implementation of the SRB is more efficient than plane-wave calculations for reduced dimensional extended systems, such as surfaces, nanorods, nanowires, and nanoribbons.
The reduction in runtime depends chiefly on the number of k-points, the band density, and the dimensionality of the BZ.
We have demonstrated speed ups compared to plane-waves in excess of 5x while maintaining 5 meV RMS accuracy across the bandstructure, 1 kbar accuracy in pressure, and $10^{-3}$ Ry$/a_0$ RMS accuracy across forces.
Reducing accuracy constraints to be in line with practical expectations for MD trajectories leads to a 1.67x performance improvement for bulk Au.

To enable highly efficient self-consistent calculations, we have described a k-independent transformation of the PWDFT Hamiltonian into the SRB and analogous transformation of the electron density to real-space.
The k-independent transformation relies on the novel auxiliary basis approach for the projectors and explicit density matrix approach for the electron density.
These techniques are examples of a broader class of k-independent transformations that could be used to efficiently transform other operators into the SRB.
For example, the treatment of electron-phonon interactions in density functional perturbation theory~\cite{Giustino2007} and the treatment of electronic excited states within many-body perturbation theory~\cite{Hybertsen1986,Rohlfing2000}, in some systems, lead to large numbers of matrix elements that need to be computed on a fine k-point grid.
Further, the k-independent transformation generalizes to any transformation of the PWDFT Hamiltonian and density matrix to another space, $\braket{b}{g}$.
It could be considered a baseline method for transformations between plane-waves and bases other than the SRB, and will be efficient as long as the other basis is small compared to plane-waves.

The SRB shifts numerical focus away from FFTs and towards dense linear algebra.
For example, the number of FFTs is independent of the number of k-points.
Dense linear algebra has a very different computational profile than FFTs.
Particularly, dense linear algebra is generally floating-point bound, while FFTs are bandwidth bound.
With the growing heterogeneity of computer architectures, this difference will have growing implications for the performance of the methods. 
If general purpose graphical processing units and other modern coprocessors are any indication, floating-point performance will continue to grow more rapidly than bandwidth, favoring dense linear algebra.

We have outlined where there is room for improvement in our current implementation of the SRB.
The most pressing issue is iterative diagonalization, which will greatly improve the performance of 2D and 3D systems, as demonstrated by a Ni slab and a finite-temperature Au snapshot here.
Additionally, the coarse meshes, q-points, recommended here are independent of the composition of the fine mesh, k-points, which is likely sub-optimal.
Pruning could provide an efficient means for updating the basis without performing additional plane-wave calculations or transformations.
When the number of k-points is large, likely in non-self-consistent calculations, dividing the BZ could reduce the subspace inefficiency.
At this early stage, with room for development in theory and in implementation, we think the SRB approach holds much promise for improving the efficiency of self-consistent electronic structure calculations.

\begin{acknowledgments}
M Hutchinson acknowledges support from the Department of Energy Computational Science Graduate Fellowship Program (Grant No. DE-FG02-97ER25308).
This work was performed at the Molecular Foundry, supported by the Office of Science, Office of Basic Energy Sciences, of the U.S. Department of Energy under Contract No. DE-AC02-05CH11231.
\end{acknowledgments}

\appendix

\section{Performance model} \label{sec:model}
Both PWDFT and SRB calculations spend the majority of their runtime in the FFT, BLAS, and LAPACK libraries.
The performance of the two methods can be well characterized by the number and size of those calls.

Consider a self-consistent calculation of an electron density using a traditional plane-wave method.
At the heart of the procedure is a matrix free iterative eigensolver, most commonly conjugate gradient (CG)~\cite{Payne1992} or Davidson~\cite{Kresse1996,Giannozzi2009}.
Both access the Hamiltonian by computing the action on a vector $H \ket{\phi}$.
We restrict the rest of this discussion to Davidson, which is considered higher performance and a good representative of related sub-space techniques.
Davidson uses the action to represent the Hamiltonian and overlap matrix in a dense subspace~\cite{Davidson1975}.
In PWDFT, each Hamiltonian evaluation requires 2 fast Fourier transforms (FFT) to access the real-space local potential, and 2 matrix-vector (MV) products to project into and out of the atomic wave function space.
The dense Hamiltonian is formed by inner-products over plane-waves.
The subspace eigenproblem is then solved directly.
Candidate solutions are transformed back into the full space and their residuals, $(H-\varepsilon S)\ket{\psi}$, are used to expand the subspace using a diagonal approximation of the inverse.
The construction of the electron density, $\rho$, requires an additional FFT and MV product per eigenfunction.
In sum, the Davidson workload per k-point per self-consistent iteration is:
\begin{multline}
W_{PW} = (2\alpha_{D} + 1)  B \left(\text{FFT}[G] + \text{MV}[P, G] \right) \\
  + MM[\alpha_D B, \alpha_D B, G] + \text{EVD}[\alpha_D B]
\end{multline}
where $B$ is the number of bands, $G$ is the number of plane-waves, $P$ is the number of projectors, $\alpha_D$ is size of the final subspace as multiple of the number of bands, and $\text{EVD}[N]$ denotes an eigenvalue decomposition (eigensolve) of size $N$ which computes all eigenvectors.
The assumption is made that the direct diagonalization of the largest subspace dominates the other diagonalizations, which underestimates the subspace diagonalization costs.
Note that in Davidson, the matrix-vector (MV) products can be coalesced into a more efficient matrix-matrix (MM) product because the subspace can be formed in parallel.
In some codes, the transformation to pseudo-wave function space is performed in real-space, in which case the MV product is sparse with a number of non-zero terms that grows linearly with the system size, not quadratically.

When using the SRB, as described in \sref{algo}, the workload becomes:
\begin{align}
W_{SRB} &= \text{SVD}[\bar{Q} B,G] + \text{MM}[\bar{Q} B, \bar{Q} B, G] + \text{MM}[\bar{Q} B, S, G] \nonumber \\
  &+ 2 S~\text{FFT}[G] + 5 \text{MM}[S, S, G] \nonumber \\
  &+ K~\text{MM}[S, P, X] + A~\text{MM}[S, X, G] \nonumber \\
  &+ K~\text{MM}[S,S,P] \nonumber \\
  &+ K~\text{EVD}[S] \\
  &+ K~\text{MM}[S, S, B] + \text{SVD}[S, S] \nonumber \\
  &+ \alpha_\rho S \left(\text{FFT}[G] + \text{MV}[S,G] \right) \nonumber \\
  &+ (2\alpha_D) Q B \left(\text{FFT}[G] + \text{MV}[P, G] \right) \nonumber
\end{align}
Where SVD denotes the singular value decomposition, $S$ is the size of the SRB, $X$ is the size of the auxiliary basis, $A$ is the number of atoms, $\bar{Q}$ is the number of q-points, $Q$ is the number of irreducible q-points, EVD denotes the eigenvalue decomposition, and $\alpha_\rho$ is the fraction of the SRB degrees of freedom needed to represent the density matrix, with a typical value of $1/4$.  

When a saved basis is used, the workload reduces to:
\begin{align}
W_{SRB'} &= S \text{FFT}[G] + \text{MM}[S, S, G] \nonumber \\
  &+ K \text{EVD}[S] \nonumber \\
  &+ K \text{MM}[S, S, B] + \text{SVD}[S, S] \\
  &+ \alpha_\rho S \left(\text{FFT}[G] + \text{MV}[S,G] \right) \nonumber \\
\end{align}
which omits the plane-wave calculation and SVD used to form the basis and the transformation of projectors and kinetic energy, which are density independent, and uses the previously computed Fourier transform of the basis elements. 

As the number of k-points increases, the cost of the EVD dominates all other SRB costs.
The Davidson solver in the plane-wave algorithm includes a similarly sized EVD per k-point.
We can write $S = \alpha_S B$ so that the ratio $\alpha_D / \alpha_S$ characterizes the relative sizes of the direct eigensolves.
In this large K regime, the speedup should be:
\begin{multline}
\text{Speedup} = \left(\frac{\alpha_D}{\alpha_S}\right)^3 
\left[1 + \frac{W_{PW} - \text{EVD}[\alpha_D B]}{\text{EVD}[\alpha_D B]} \right]
\end{multline}
When $\alpha_D / \alpha_S < 1/2$, the performance of the reduced diagonalization degrades rapidly.
In these cases, $\alpha_S >> 1$, so most of the direct solver's effort is wasted on uninteresting eigenpairs.
An iterative eigensolver should be used instead.

As the number of plane-waves per SRB matrix element, $G / S$, increases, the cost of the G-dependent terms dominates all other costs.
The Fourier transforms are the highest order in $G$, so we only count them.
The plane-wave calculation uses $ 2 K B \alpha_D$ while the SRB uses $2 Q B \alpha_D + 2 S = 2 B (Q \alpha_D + \alpha_S)$.
The speed-up should approach
\begin{equation} \label{eqn:klimit}
\text{Speedup} = \frac{K \alpha_D}{Q \alpha_D + \alpha_S}
\end{equation}
Because $G$ grows linearly with the volume but $S$ grows linearly with the number of bands, $B$, the ratio $S/G \approx B/V$, the \textit{band density}.
The lower the band density, the better one can expect the SRB to perform.

\bibliography{library}

\end{document}